\documentclass[journal,11pt,draftclsnofoot,onecolumn]{IEEEtran}

\ifCLASSINFOpdf
\usepackage[pdftex]{graphicx}
  \graphicspath{{../pdf/}{../jpeg/}}
  \DeclareGraphicsExtensions{.pdf,.jpeg,.png}
\else
  \usepackage[dvips]{graphicx}
  \graphicspath{{../eps/}}
  \DeclareGraphicsExtensions{-eps-converted-to.pdf}
\fi
\usepackage{epstopdf}
\usepackage{amsmath,amsthm, amssymb}
\usepackage{threeparttable}
\usepackage{bm}
\usepackage{multirow}
\usepackage{booktabs}
\usepackage{color}
\usepackage{tabularx}
\usepackage{setspace}
\usepackage{verbatim}
\usepackage{stfloats}
\usepackage{caption}
\usepackage{float}
\usepackage[T1]{fontenc}
\usepackage{soul}
\usepackage{cite}

\usepackage{stfloats}
\usepackage{mathtools}
\usepackage{amsmath}
\allowdisplaybreaks[0]
\usepackage{subfigure}
\usepackage{fancyhdr}
\usepackage{cancel}

\makeatletter
\newif\if@restonecol
\makeatother

\usepackage[linesnumbered,ruled,vlined]{algorithm2e}
\usepackage{algpseudocode}
\usepackage{amsmath}

\usepackage{lineno}

\begin{document}

\title{Decentralized Power Allocation for MIMO-NOMA Vehicular Edge Computing Based on Deep Reinforcement Learning}

\author{Hongbiao Zhu, Qiong Wu, ~\IEEEmembership{Member,~IEEE}, Xiao-Jun Wu, ~\IEEEmembership{Member,~IEEE}, Qiang Fan, \\Pingyi Fan, ~\IEEEmembership{Senior Member,~IEEE}, and Jiangzhou Wang, ~\IEEEmembership{Fellow,~IEEE}

\thanks{
This work was supported in part by the National Natural Science Foundation of China under Grant No. 61701197, in part by the Beijing Natural Science Foundation under Grant No. 4202030, in part by the 111 Project under Grant No. B12018. \emph{(Corresponding author: Qiong Wu.)}

Hongbiao Zhu and Qiong Wu are with the School of Internet of Things Engineering, Jiangnan University, Wuxi 214122, China (e-mail: hongbiaozhu@stu.jiangnan.edu.cn, qiongwu@jiangnan.edu.cn).

Xiao-Jun Wu is with the School of Artificial Intelligence and Computer Science, Jiangnan University, Wuxi 214122, China (wu\_xiaojun@jiangnan.edu.cn).

Qiang Fan is with Wistron AiEdge, San Jose, CA 95131, USA (e-mail: qiang\_fan@wistron.com).

Pingyi Fan is with the Department of Electronic Engineering, Beijing National Research Center for Information Science and Technology, Tsinghua University, Beijing 100084, China (email: fpy@tsinghua.edu.cn).

Jiangzhou Wang is with the School of Engineering, University of Kent, CT2 7NT Canterbury, U.K. (Email: j.z.wang@kent.ac.uk).

}
}

\markboth{IEEE Internet of Things Journal, ~Vol.~XX, No.~XX, XXX~2021}
{Zhu \MakeLowercase{\textit{et al.}}: Decentralized Power Allocation for MIMO-NOMA Vehicular Edge Computing Based on Deep Reinforcement Learning}

\maketitle

\begin{abstract}
Vehicular edge computing (VEC) is envisioned as a promising approach to process the explosive computation tasks of vehicular user (VU). In the VEC system, each VU allocates power to process partial tasks through offloading and the remaining tasks through local execution. During the offloading, each VU adopts the multi-input multi-out and non-orthogonal multiple access (MIMO-NOMA) channel to improve the channel spectrum efficiency and capacity. However, the channel condition is uncertain due to the channel interference among VUs caused by the MIMO-NOMA channel and the time-varying path-loss caused by the mobility of each VU. In addition, the task arrival of each VU is stochastic in the real world. The stochastic task arrival and uncertain channel condition affect greatly on the power consumption and latency of tasks for each VU. It is critical to design an optimal power allocation scheme considering the stochastic task arrival and  channel variation to optimize the long-term reward including the power consumption and latency in the MIMO-NOMA VEC. Different from the traditional centralized deep reinforcement learning (DRL)-based scheme, this paper constructs a decentralized DRL framework to formulate the power allocation optimization problem, where the local observations are selected as the state. The deep deterministic policy gradient (DDPG) algorithm is adopted to learn the optimal power allocation scheme based on the decentralized DRL framework. Simulation results demonstrate that our proposed power allocation scheme outperforms the existing schemes.
\end{abstract}


\IEEEpeerreviewmaketitle

\section{Introduction}
\label{sec1}
\IEEEPARstart{W}{ith} the increasing number of vehicles, the growing demand of computation-intensive applications such as virtual/augmented reality (VR/AR), image processing, face detection and recognition is emerging to satisfy the infotainment experience of vehicular users (VUs) \cite{Bonadio2020}. These applications are realized through collecting a great amount of data by various vehicular user equipments such as smart phones and wearable devices. Such large amount of data results in intensive computation tasks which need to be processed in time, thus leading to heavy computation burden for VUs \cite{Wu2019,Hongmei}. Vehicular edge computing (VEC) is a promising way to relieve the burden\cite{Wang2017}, where a VEC server with high computational capability is connected with a base station (BS) to provide VUs with computation resources at the edge \cite{VEC}, \cite{Hou2020}. When a VU has some tasks to process, it can either offload the tasks to the VEC server collocated connecting with the BS \cite{2018Velocity}, or execute the tasks locally. For the task offloading, the VU has to consume energy in the data transmission, where the offloading power is defined as the transmission power. In addition, when the VU processes the tasks locally, the local task processing will incur the energy consumption at its central processing unit (CPU). For simplicity, we define the power consumption of task processing at the VU as the local execution power.


During the offloading, the multi-input multi-out and non-orthogonal multiple access (MIMO-NOMA) channel is considered here due to its high channel spectrum efficiency and channel capacity. Specifically, each VU can share the whole spectrum and undivided bandwidth to offload tasks and the BS is equipped with multi-antenna to receive tasks from all VUs simultaneously \cite{Liu2020,Qian2020,Marzetta,Di2017}. However, the channel condition is time varying due to the channel interference among VUs caused by the MIMO-NOMA channel and the time-varying path-loss caused by the mobility of each VU. In addition, the task arrival of each VU is stochastic in practice. The stochastic task arrival and uncertainty of channel condition significantly impact the power consumption and latency of task processing for each VU. For example, a VU would take more time in task offloading when task arrival rate is increasing and the channel condition is getting deteriorated, which increases the power consumption and latency. In this case, the VU should allocate more local execution power to reduce the power consumption and latency. In the VEC, vehicular user equipment has limited energy and the applications such as VR/AR and real-time interactive 3D gaming should be processed within a limited time, therefore power consumption and latency are two important performance metrics in task processing \cite{Ge2020,Zheng2015,Wan2021}. It is critical to design an optimal power allocation scheme considering the stochastic task arrival and uncertainty of channel condition in the MIMO-NOMA VEC.

Deep reinforcement learning (DRL) is a favorable framework to formulate the similar optimization problem in complex environments \cite{DRL}. Many existing works have designed the offloading scheme based on the centralized DRL framework in VEC by taking various factors into account, where the BS first collects the global information including all VUs' states to determine the action of each VU, which causes huge overhead and extra  latency \cite{Zhan2019,zhan2020,HWang2020,Dong2020,Ke2020,He2018,Tan2018,Ning2020,Liu2019,Luo2020,Qiao2020,Ren2020}. Only a few works focused on decentralized DRL-based offloading schemes, where each VU collects the local observations to select its action, thus the overhead and latency can be reduced efficiently\cite{Ye2019,Xu2020}. However, these works did not consider the channel caused by employing the MIMO-NOMA mode. To the best of our knowledge, no work has considered the stochastic task arrival and the uncertainty of MIMO-NOMA channel condition in the decentralized DRL-based optimal power allocation scheme in VEC.

In this paper, we consider the stochastic task arrival, and the channel condition uncertainty caused by the MIMO-NOMA channel interference and the mobility of VUs, and propose a decentralized DRL-based power allocation scheme to optimize the long-term reward in VEC in terms of power consumption and latency. The main contributions of this paper are summarized as follows.

\begin{itemize}

\item[1)] We formulate the power allocation optimization problem, where the state, action and reward function are elaborately defined to enable each VU to learn optimal power allocation scheme according to the local observations. Then, the deep deterministic policy gradient (DDPG) algorithm is adopted to learn the optimal power allocation decision based on the DRL framework.


\item[2)] Extensive experiments are carried out to test the performance of the proposed scheme and show its superiority to other existing polices in terms of power consumption and latency of task processing.

\end{itemize}

The rest of the paper is organized as follows. Section \ref{sec2} reviews the related work. Section \ref{sec3} introduces the system model. In Section \ref{sec4} the decentralized DRL framework is set up to formulate the power allocation problem. Section \ref{sec5} presents the DDPG algorithm on how to learn the optimal power allocation scheme based on the DRL framework. Section \ref{sec6} presents the simulation results. It is concluded in Section \ref{sec7}.

\section{Related Work}
\label{sec2}
In this section, we first review the related works on the offloading scheme in mobile edge computing (MEC) considering the MIMO or NOMA channel, then we review the existing works on the DRL-based offloading scheme in the VEC.

\subsection{Offloading in MIMO or NOMA MEC}
In recent years, many works have considered the MIMO or NOMA channel  while designing the offloading scheme in MEC.

In \cite{Wang2017}, Wang \emph{et al.} employed NOMA channel in MEC computation offloading system to minimize the energy consumption of all users where Lagrange dual was adopted to make decisions about task offloading proportion, successive interference cancellation order, offloading power and local CPU frequencies.
In \cite{Pan2019}, Pan \emph{et al.} considered NOMA channel in MEC for uploading computation tasks and downloading computation result where convex optimization was adopted to minimize the energy consumption by determining offloading task partitions, offloading power and task time allocation.
In \cite{Huang2019}, Huang \emph{et al.} focused on the channel estimation process with pilots in massive MIMO MEC system to minimize the offloading latency of all users by optimizing power of pilot transmission and data transmission, as well as the allocation of computing resource.
In \cite{Ding2021}, Ding \emph{et al.} studied a multi-user MIMO (MU-MIMO) MEC system to minimize the system cost, the weighted sum of latency and energy consumption.
In \cite{Feng2020}, Feng \emph{et al.} considered the fairness of all users in a MU-MIMO MEC system and to minimize offloading latency through optimizing the distribution of resource, transmission of pilot sequence and data.
However, these works did not consider the scenario of vehicular scenarios.

\subsection{DRL-based Offloading in VEC}
Many works have discussed DRL-based offloading scheme in VEC.
In \cite{Dong2020}, Dong \emph{et al.} considered NOMA channel in VEC where Deep Q-Network (DQN) was applied to guarantee the delay requirement and minimize the energy consumption.
In \cite{Ke2020}, Ke \emph{et al.} designed a three-layer VEC offloading system including a macro BS, multiple small BSs and vehicles where DRL is applied to minimize the cost consisting of energy consumption and transmission delay.
In \cite{He2018}, He \emph{et al.} proposed an offloading scheme considering network, cache, and computation resource in VEC, where DQN was employed to select the optimal offloading decision that maximizes the reward including the caching state, computation capability and received signal-to-noise ratio (SNR).
In \cite{Tan2018}, Tan \emph{et al.} formulated the joint optimal caching and computing allocation problem to minimize the VEC system cost including communication, computation and storage under the constraint of server deadline. DQN was employed to solve the optimization problem where the channel was assigned by orthogonal frequency division multiplexing (OFDM).
In \cite{Ning2020}, Ning \emph{et al.} constructed an edge computation and cache model for VEC consisting of macro BS, several RSUs and VUs, where the tasks of VUs were divided into computing tasks and content tasks. DDPG was employed to obtain the optimal resource allocation in order to maximize the reward of mobile network operator (MNO), including computing and caching cost, penalty on quality of experience (QoE).
In \cite{Liu2019}, Liu \emph{et al.} took stochastic traffic and uncertain communication conditions of VEC into consideration, and adopted the semi-Markov process to formulate an optimization problem to maximize the total network utility of the VEC. The DQN method was employed to obtain the optimal offloading scheme.
In \cite{Ren2020}, Ren \emph{et al.} designed a VEC architecture consisting of BSs, RSUs, VUs, and cell software designed network controller where tasks can be migrated among BSs and RSUs. A centralized DRL-based offloading scheme was designed to manage the network resource through making decisions of offloading, migration and resource allocation.
However, these works only focused on the centralized DRL-based offloading scheme.

A few works have also focused on decentralized DRL-based offloading schemes in VEC \cite{Ye2019,Xu2020}.
In \cite{Ye2019}, Ye \emph{et al.} considered a VEC that is composed of vehicle-to-vehicle (V2V) and vehicle-to-infrastructure (V2I) communications where V2I communication reserved orthogonal single-input single-output (SISO) channels. DQN was adopted to select the task transmitting sub-band and power level for VUs to maximize the reward consisting of system communication capacity and latency.
In \cite{Xu2020}, Xu \emph{et al.} considered the similar scenario \cite{Ye2019} to maximize the sum-rate of every sub-band communication where DDPG was employed to obtain the optimal policy in continuous action space. However, \cite{Xu2020,Ye2019} did not consider the channel varying caused by the MIMO-NOMA channel interference and the mobility of VU in VEC.

As mentioned above, no work has considered the stochastic task arrival and channel varying in the MIMO-NOMA VEC while designing the decentralized DRL-based power allocation scheme.

\begin{figure*}
\centering
\includegraphics[width=5.5in]{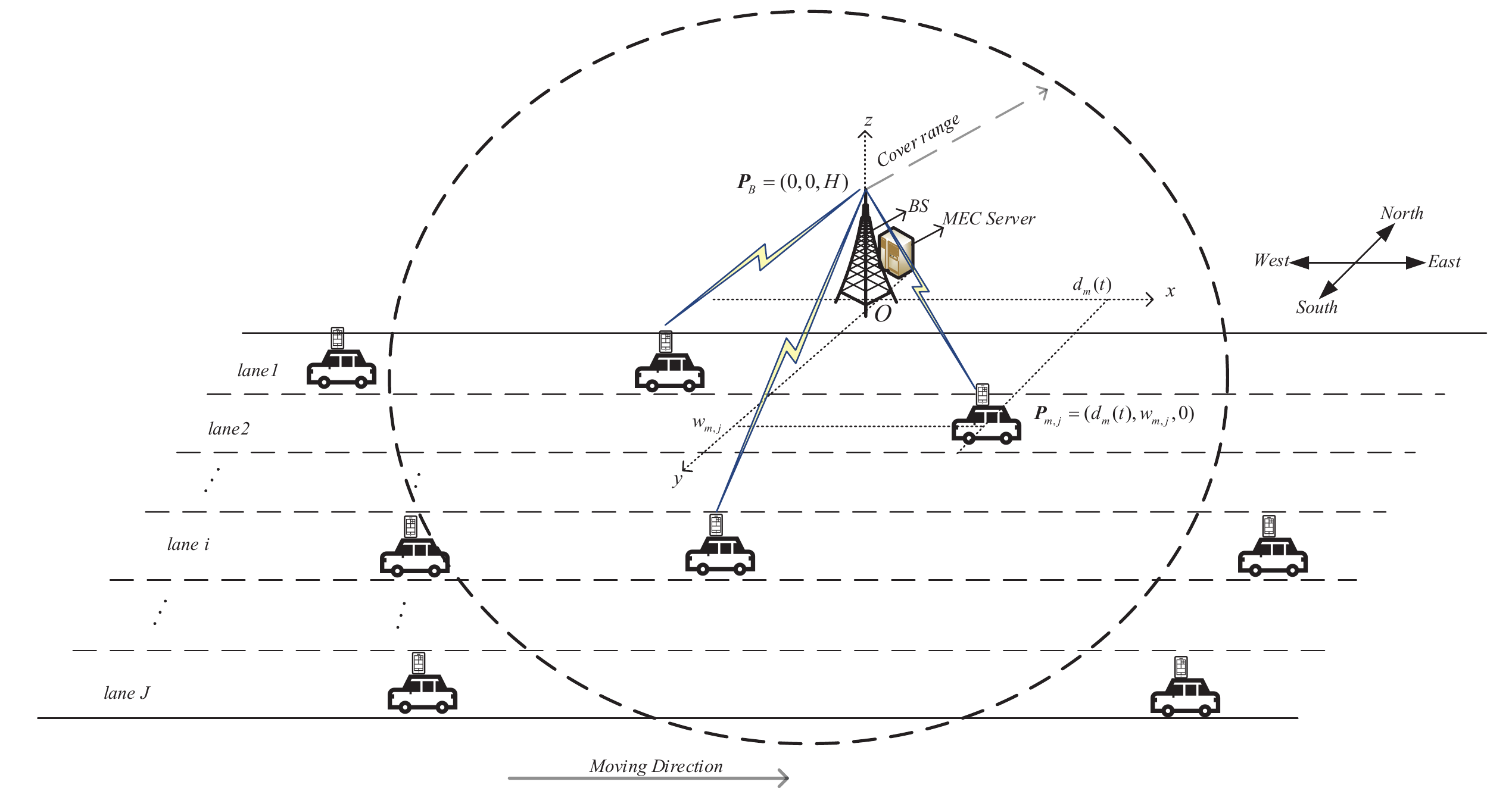}
\caption{System model.}
\label{fig1}
\end{figure*}

\section{System Model}
\label{sec3}

\begin{table*}
\footnotesize
\caption{The summary for notations.}
\label{tab1}
\centering
\begin{tabular}{|p{2.1cm}<{\centering}|p{5.8cm}|p{2cm}<{\centering}|p{5.8cm}|}
\hline
\textbf{Notation} &\textbf{Description} &\textbf{Notation} &\textbf{Description}\\
\hline
$a_m(t)$ &Task bits of user $m$ arrived at slot $t$. &$a_{m,t}$ &Action space of VU $m$ at slot $t$.\\
\hline
$a_m$ &Abbreviation of $a_{m,t}$. &$a_{m}'$ &Abbreviation of $a_{m,t+1}$.\\
\hline
$a_m^i$ &Action of the $i$-th tuple in mini-batch. &$B_m(t)$ &Buffer length of VU $m$ at slot $t$.\\
\hline
$\mathcal{R}$ &Replay buffer. &$D$ &Diameter of BS's coverage.\\
\hline
$d_{m,l}(t)$ &Bits of user $m$ processed locally at slot $t$. &$d_{m,o}(t)$ &Bits of user $m$ offload at slot $t$.\\
\hline
$d_m(t)$ &\multicolumn{1}{m{6cm}|}{Distance between user $m$ and BS's antennas along the $x$-axis.} &$\boldsymbol{e}(t)$ &Error vector of adjacent slots channel vector.\\
\hline
$f_m(t)$ &CPU frequence of user $m$ at slot $t$. &$F_{max}$ &The maximum allowed CPU frequence. \\
\hline
$\boldsymbol{h}_{m}^{s}(t)$ &\multicolumn{1}{m{6cm}|}{Small-scale Rayleigh fading channel gain of VU $m$ at slot $t$.} &$h_{m}^p(t)$ &\multicolumn{1}{m{6cm}|}{The large-scale fading coefficient reflects the path-loss of VU $m$ at slot $t$.}\\
\hline
$h_r$ &Reference power gain at distance $1$m. &$\boldsymbol{H}(t)$ &\multicolumn{1}{m{6cm}|}{Channel matrix between BS and every VU at slot $t$.}\\
\hline
$H$ &Height of BS. &$I$ &Size of mini-batch. \\
\hline
$K_{max}$ &The maximum episode in training stage. &$i$ &Index of tuples in the mini-batch.\\
\hline
$J(\mu_m)$ &Objective function. &$L_m$ &CPU cycles needed for VU $m$ process one bit.\\
\hline
$L(\zeta^m)$ &Loss function. &$N_j$ &Total number of time slots.\\
\hline
$\boldsymbol{n}(t)$ &Noise of signal received by BS. &$N$ &Number of antenna.\\
\hline
$N_{max}$ &The maximum number of VUs in the sytem. &$\boldsymbol{P}_{m,j}(t)$ &\multicolumn{1}{m{6cm}|}{Location of VU $m$ at slot $t$.} \\
\hline
$\boldsymbol{P}_B$ &Location of  BS. &$p_{m,o}(t)$ &Offload power offered by VU $m$ at slot $t$.\\
\hline
$p_{m,l}(t)$ &\multicolumn{1}{m{6cm}|}{Local process power offered by VU $m$ at slot $t$.} &$P_{max,o}$ &Maximum offload power. \\
\hline
$P_{max,l}$ &Maximum local execution power. &{$Q^{\mu_{\theta^m}}(s_{m,t},a_{m,t})$}&\multicolumn{1}{m{6cm}|}{Action-value function of VU $m$ following policy $\mu_{\theta^m}$.} \\
\hline
$Q^{\zeta^m}(s_{m,t},a_{m,t})$ &\multicolumn{1}{m{6cm}|}{Action-value function approximated by critic-network.} &$Q^{\zeta^{m'}}(s_{m,t},a_{m,t})$ &\multicolumn{1}{m{6cm}|}{Action-value function of VU $m$ approximated by target critic-network.}\\
\hline
$r_{m,t}$ &Reward of VU $m$ slot $t$. &$r_m^i$ &Reward of the $i$-th tuple in mini-batch.\\
\hline
$r_{m}$ &Abbreviation of $r_{m,t}$. &$\mathcal{R}$ &Replay buffer.\\
\hline
$s_{m,t}$ &State space of VU $m$ lot $t$. &$s_m$ &Abbreviation of $s_{m,t}$.\\
\hline
$s_m'$ &Abbreviation of $s_{m,t+1}$ &$s_m^i$ &State of the $i$-th tuple in mini-batch. \\
\hline
${s'}_{m}^{i}$ &Next state of the $i$-th tuple in mini-batch. &$T_s$ &Safety time.\\
\hline
$v_j$ &\multicolumn{1}{m{6cm}|}{Velocities of vehicles driven on lane $j$.} &$v_m$ &velocity of vehicular user $m$.\\
\hline
$w_0$ &\multicolumn{1}{m{6cm}|}{The width between the VU driven on the lane $1$ and BS's antennas along $y$-axis.} &$w$ &Width of roads. \\
\hline
$w_{m,j}$ &\multicolumn{1}{m{6cm}|}{Width between VU $m$ driven on lane $j$ and antennas along the $y$-axis.} &$W$ &Bandwidth. \\
\hline
$\boldsymbol{y}(t)$ &Signal received by BS. &$y_{m}^{i}$ &Target value.\\
\hline
$\gamma$ &Discounting factor of long-term reward. &$\gamma_m(t)$ &SINR of VU $m$ at slot $t$. \\
\hline
$\Delta_t$ &The exploration noise at slot $t$. &$\zeta^m$ &Parameter of critic network.\\
\hline
$\zeta^{m^{\prime}}$ &Parameter of target critic-network. &$\eta$ &path-loss exponent.\\
\hline
$\theta^{m}$ &Parameter of actor network. &$\theta^{m*}$ &Optimized parameter of actor network.\\
\hline
$\theta^{m^{\prime}}$ &Parameter of target actor-network. &$\kappa_m$ &Effective switched capacitance of user $m$.\\
\hline
$\lambda_m$ &Mean rate of tasks arrival for VU $m$. &$\mu_{\theta^m}$ &Policy of VU $m$ approximated by actor network.\\
\hline
$\rho_m$ &\multicolumn{1}{m{6cm}|}{Normalized channel correlation coefficient of user $m$ between adjacent slots.} &$\sigma_{R}^{2}$ &\multicolumn{1}{m{6cm}|}{Additive white Gaussian noise variance of the signal received by BS.}\\
\hline
$\tau$ &Update parameter for target networks. &$\tau_0$ &Slot duration.\\
\hline
$\omega_{1},\omega_{2}$ &Weighted factors of reward.\\
\hline
\end{tabular}
\end{table*}

The system model is shown as Fig. \ref{fig1}. Consider a VEC system where an $N$-antenna BS is placed along a one-way $J$-lane road and a VEC server is attached to the BS. The lanes from near to far according to the vertical distance to the BS are denoted as $1$, $2$, $\cdots$, $j$, $\cdots$, $J$, respectively. $M$ vehicles on different lanes traverse the coverage of BS from left to right with different velocities, where each vehicle carries a computation resource-limited single-antenna VU. The duration time that a VU on lane $j$ stays in the transmission coverage of BS is divided into $N_j$ equal time slots, each of which is a constant $\tau_0$.  At each slot, computation-intensive tasks arrive at the first come first service (FCFS) buffer of each VU stochastically following independent and identical distribution (i.i.d.). Meanwhile, each VU allocates the local execution power and offloading power to process the tasks stored in the buffer queue locally or at the VEC server nearby, respectively. Moreover, the channel condition varies due to the interference among VUs' MIMO-NOMA channel and the time-varying path-loss caused by the mobility of VUs. During the offloading, each VU first transmits tasks to the BS, then the BS processes the tasks and adopts the zero-forcing (ZF) technique to detect the received signal and noise of each VU from the received signal of all VUs and further determines the signal-to-interference-plus-noise ratio (SINR) of each VU. Afterwards the BS sends back the computation results as well as the determined SINR to each VU at the next slot.  Different from the traditional centralized DRL-based offloading scheme in VEC, in this paper each VU can distributively determine the power allocation based on its local information. Next, we will introduce the computation model, network model and mobility model to formulate the local information of VU $m$ such as the buffer length, SINR and position, respectively. For simplicity, the notations adopted in this paper are listed as TABLE \ref{tab1}.

\subsection{Mobility model}

Let $\boldsymbol{P}_{m,j}(t)$ be the position of VU $m$ which moves on lane $j$ at slot $t$. A space rectangular coordinate system shown in Fig. \ref{fig1} is set to illustrate the positions of each VU and the BS, where the origin is the position of the BS, the direction of the $x$-axis is the moving direction of VUs, i.e., east, the direction of the $y$-axis is south, the direction of $z$-axis is set along the antennas of BS which is perpendicular to both $x$-axis and $y$-axis. Let $d_m(t)$ and $w_{m,j}$ be the distances between VU $m$ and the antennas of BS along $x$-axis and $y$-axis at slot $t$, respectively. Thus $\boldsymbol{P}_{m,j}(t)$ is denoted as $(d_m(t),w_{m,j},0)$ in the space rectangular coordinate system, where $w_{m,j}$ depends on the lane index of VU $m$, i.e., $j$, and is calculated as

\begin{equation}
 w_{m,j}= (j-1)\cdot w+w_0,
\label{eq1}
\end{equation}
here $w$ is the width of a lane and $w_0$ is the distance between lane $1$ and the BS's antennas along $y$-axis.

Similar to \cite{distance}, the position of VU $m$ is approximately constant within each time slot due to the sufficiently small value of $\tau_0$. Since VU $m$ moves on lane $j$ with a constant velocity $v_j$, $d_m(t)$ is updated as
\begin{equation}
 d_m(t)=d_m(t-1)+v_j\tau_0, d_m(t) \in [-\frac{D}{2},\frac{D}{2}],
\label{eq2}
\end{equation}
where $D$ is the coverage of the BS and $d_m(1) = -\frac{D}{2}$. VU $m$ communicates with the BS once it enters the coverage of the BS and calculates $d_m(t)$ at each slot $t$ according to Eq. \eqref{eq2}. Therefore, $d_m(t)$ is a local observation of VU $m$ at slot $t$ to reflect the mobility of VU $m$.

Note that according to the 4-second rule \cite{foursrule}, the maximum number of VUs on lane $j$ can be calculated as $\lfloor D/(v_j\cdot T_s)\rfloor$, where $T_s$ is the safety time, i.e., $4$ $seconds$. Thus, the maximum number of VUs in the sytem can be calculated as 

\begin{equation}
N_{max}=\sum^{J}_{j=1}\lfloor D/(v_j\cdot T_s)\rfloor.
\label{eqNmax}
\end{equation}

\subsection{Network model}

The channel matrix at slot $t$ can be expressed as $\boldsymbol{H}(t)=[\boldsymbol{h}_1(t), \cdots, \boldsymbol{h}_m(t), \cdots, \boldsymbol{h}_M(t)]\in \mathbb{C}^{N\times M}$, where $\boldsymbol{h}_m(t)\in\mathbb{C}^{N\times 1}$ is the channel vector between VU $m$ and BS. In the MIMO-NOMA channel, the signal received by BS at slot $t$ is the signal transmitted from all VUs, which can be expressed as
\begin{equation}
\begin{aligned}
\boldsymbol{y}(t)=\sum_{m\in\mathcal{M}}\boldsymbol{h}_m(t)\sqrt{p_{m,o}(t)}s_m(t)+\boldsymbol{n}(t), \\ p_{m,o}(t) \in {[0,P_{max,o}]},
\label{eq3}
\end{aligned}
\end{equation}
where $p_{m,o}(t)$ is the offloading power of VU $m$ at $t$, $P_{max,o}$ is the maximum offloading power, $s_m(t)$ is the complex data symbol with unit variance, and $\boldsymbol{n}(t)$ is the vector of additive white Gaussian noise (AWGN) with variance $\sigma_{R}^{2}$, (i.e., $\boldsymbol{n}(t)\sim \mathcal{CN}(\boldsymbol{0},\sigma_{R}^{2}\boldsymbol{I}_N)$, and $\boldsymbol{I}_N$ is an $N \times N$ identity vector). Furthermore, $\boldsymbol{h}_m(t)$ is an integrated one of the stochastic small-scale fading channel gain $\boldsymbol{h}_{m}^s(t)$ and the large-scale fading coefficient ${h}_{m}^p(t)$ which reflects the path-loss of VU $m$ \cite{zhan2020}, i.e.,

\begin{equation}
\boldsymbol{h}_{m}(t)=\boldsymbol{h}_{m}^s(t)\sqrt{h_{m}^p(t)}.
\label{eq4}
\end{equation}
In Eq. \eqref{eq4}, $h_{m}^p(t)$ characterizes the mobility of VU $m$ and is calculated as
\begin{equation}
h_{m}^p(t)=\frac{h_{r}}{\left\|{\boldsymbol{P}_{m,j}(t)}-{\boldsymbol{P}_{B}}\right\|^{\eta}},
\label{eq5}
\end{equation}
where $\eta$ is the path-loss exponent, $h_r$ is the channel power gain at $1$ meter distance, $\boldsymbol{P}_{m,j}(t)=(d_m(t),w_{m,j},0)$ is the position of VU $m$ at slot $t$ and $\boldsymbol{P}_B$ is the position of antennas of BS. Note that $w_{m,j}$ is calculated according to Eq. \eqref{eq1} and $d_m(t)$ is calculated according to Eq. \eqref{eq2}. Let $H$ be the height of antennas of BS, thus $\boldsymbol{P}_B=(0,0,H)$.


In addition, the following autoregressive (AR) model is adopted to formulate the relationship between $\boldsymbol{h}_{m}^{s}(t)$ and $\boldsymbol{h}_{m}^{s}(t-1)$ \cite{ZF}, i.e.,
\begin{equation}
\boldsymbol{h}_{m}^s(t)=\rho_m\boldsymbol{h}_{m}^s(t-1)+\sqrt{1-\rho_{m}^2}\boldsymbol{e}(t),
\label{eq6}
\end{equation}
where $\rho_m$ is the normalized channel correlation coefficient between the consecutive slots, $\boldsymbol{e}(t)$ is the error vector which obeys the complex Gaussian distribution and is correlated with $\boldsymbol{h}_{m}^{s}(t)$.

According to Jake's fading spectrum, $\rho_m =J_{0}(2\pi f_{d}^m\tau_0)$, where $J_{0}(\cdot)$ is the zeroth-order Bessel function of the first kind and $f_d^m$ is the Doppler frequency of VU $m$\cite{Doppler}, which can be calculated as

\begin{equation}
f_d^m = \frac{v_m}{\Lambda} \cos\Theta,
\label{doppler}
\end{equation}
where $\Lambda$ is the wavelength, and $\Theta$ is the angle between moving direction, i.e., $\boldsymbol{x}_0=(1,0,0)$, and uplink communication direction, i.e., $\boldsymbol{P}_{B} - \boldsymbol{P}_{m,j}(t)$. Thus $\cos \Theta$ can be calculated as

\begin{equation}
\cos \Theta = \frac{\boldsymbol{x}_0 \cdot (\boldsymbol{P}_{B} - \boldsymbol{P}_{m,j}(t))}{\left\|{\boldsymbol{P}_{B}-\boldsymbol{P}_{m,j}(t)}\right\|}.
\label{cos}
\end{equation}

Then the BS adopts the pseudo inverse of $\boldsymbol{H}(t)$, which is denoted as $\boldsymbol{H}^{\dagger}(t)$, as the ZF detector to detect the received signal of VU $m$ from $\boldsymbol{y}(t)$. According to \cite{ZF}, $\boldsymbol{H}^{\dagger}(t)$ is calculated as

\begin{equation}
\boldsymbol{H}^{\dagger}(t)=\left(\boldsymbol{H}^{H}(t)\boldsymbol{H}(t)\right)^{-1}\boldsymbol{H}^{H}(t),
\label{eq7}
\end{equation}
where $\boldsymbol{H}^{H}(t)$ is the conjugate transpose of $\boldsymbol{H}(t)$.

Specifically, letting $\boldsymbol{g}_{m}^{H}(t)$ be the $m$-th row of $\boldsymbol{H}^{\dagger}(t)$, by multiplying $\boldsymbol{y}(t)$ with $\boldsymbol{g}_{m}^{H}(t)$, we can obtain the following equation according Eq. \eqref{eq3}:

\begin{equation}
\boldsymbol{g}_{m}^{H}(t)\boldsymbol{y}(t)=\boldsymbol{g}_{m}^{H}(t)\sum_{m\in\mathcal{M}}\boldsymbol{h}_m(t)\sqrt{p_{m,o}(t)}s_m(t)+\boldsymbol{g}_{m}^{H}(t)\boldsymbol{n}(t).
\label{eq8}
\end{equation}

Since $\boldsymbol{g}_{m}^{H}(t)$ is the $m$-th row of $\boldsymbol{H}^{\dagger}(t)$, according to Eq. \eqref{eq7}, we have
\begin{spacing}{1.35}
\begin{equation}
\boldsymbol{g}_{m}^{H}(t)\boldsymbol{h}_{i}(t)=\delta_{m,i}(t)=\\
\left\{
\begin{array}{l}
{1, \qquad i=m}\\
{0, \qquad i\neq m}
\end{array}
\right.
\label{eq9}
\end{equation}
\end{spacing}

Substituting Eq. \eqref{eq9} into Eq. \eqref{eq8}, we have
\begin{equation}
\boldsymbol{g}_{m}^{H}(t)\boldsymbol{y}(t)=\sqrt{p_{m,o}(t)}s_m(t)+\boldsymbol{g}_{m}^{H}(t)\boldsymbol{n}(t),
\label{eq10}
\end{equation}
where $\sqrt{p_{m,o}(t)}s_m(t)$ is the signal received by BS from VU $m$, $\sqrt{p_{m,o}(t)}$ is the power of the received signal, and $\boldsymbol{g}_{m}^{H}(t)\boldsymbol{n}(t)$ is the noise of VU $m$ received by BS. Since the power of $\boldsymbol{n}(t)$ is $\sigma_{R}^{2}$, the noise power is calculated as $\left\|{\boldsymbol{g}_m^H(t)}\right\|^2 \sigma_{R}^{2}$. Thus, the SINR of VU $m$ can be calculated as

\begin{equation}
\gamma_m(t)=\frac{p_{m,o}(t)}{\left\|{\boldsymbol{g}_m^H(t)}\right\|^2 \sigma_{R}^{2}}.
\label{eq11}
\end{equation}

The BS can detect the SINR of VU $m$ at slot $t$, i.e., $\gamma_m(t)$, according to Eqs. \eqref{eq3}-\eqref{eq11} and transmit $\gamma_m(t)$ to VU $m$ at next slot. In this case, VU $m$ receives $\gamma_m(t-1)$ at slot $t$ and thus $\gamma_m(t-1)$ is also a local observation of VU $m$ at slot $t$ to reflect the channel variation.

\subsection{Computation model}

For VU $m$, the buffer length of VU $m$ at slot $t$ is denoted as $B_m(t)$ and the relationship between $B_m(t)$ and $B_m(t-1)$ is expressed as

\begin{equation}
{B}_m(t)=[B_m(t-1)-(d_{m,o}(t-1)+d_{m,l}(t-1))]^{+}+a_m(t-1),
\label{eq12}
\end{equation}
where $[\cdot]^{+}=\max(0,\cdot)$, $a_m(t-1)$ is the amount of the tasks arriving at the buffer queue of VU $m$ at slot $t-1$, $d_{m,l}(t-1)$ and $d_{m,o}(t-1)$ are the amount of the tasks processed by local execution and offloaded to the BS at slot $t-1$, respectively. Therefore the amount of tasks departing from the buffer at slot $t-1$ becomes $d_{m,l}(t-1) + d_{m,o}(t-1)$, which should not exceed $B_m(t-1)$. We will also explain how $d_{m,l}(t-1)$ and $d_{m,o}(t-1)$ are determined as follows.

\subsubsection{Local execution}

Let $L_m$ be the computation intensity of tasks (i.e., the number of CPU cycles required to processed one bit data), $f_m(t-1)$ be the CPU frequency of VU $m$ at slot $t-1$. The task size that can be processed by local execution at slot $t-1$ is calculated as

\begin{equation}
d_{m,l}(t-1)=\tau_0f_m(t-1)/L_m.
\label{eq13}
\end{equation}
Letting $p_{m,l}(t-1)$ be the local execution power at slot $t-1$, $f_m(t-1)$ is calculated as 

\begin{equation}
\begin{aligned}
f_m(t-1)&=\sqrt[3]{p_{m,l}(t-1)/\kappa}, \\
& p_{m,l}\in[0,P_{max,l}], f_m(t-1)\in[0,F_{max}],
\label{eq14}
\end{aligned}
\end{equation}
where $\kappa$ is the effective switched capacitance, $P_{max,l}$ is the maximum local execution power, $F_{max}$ is the maximum CPU frequency. According to Eq. \eqref{eq14}, $F_{max}$ can be calculated as $F_{max}=\sqrt[3]{P_{max,l}/\kappa}$.

\subsubsection{Offloading}

In the offloading mode, the computation resources of VEC server are sufficient, thus the latency that the VEC server processes the tasks is negligible. In addition, the size of computation result is usually very small, thus the feedback delay can also be ignored. Therefore, the delay of task transmission is the duration of a slot $\tau_0$. In this case, the amount of tasks processed by offloading at slot $t-1$ can be calculated according to Shannon theory, i.e.,
\begin{equation}
d_{m,o}(t-1)=\tau_0W\log_2(1+\gamma_m(t-1)),
\label{eq15}
\end{equation}
where $W$ is the bandwidth and $\gamma_m(t-1)$ is the SINR of VU $m$ at slot $t-1$.

Since VU $m$ receives its SINR information $\gamma_m(t-1)$ from the BS at slot $t$, it allocates $p_{m,l}(t-1)$ and observes $a_m(t-1)$ at slot $t-1$. In this case, VU $m$ can calculate $B_m(t)$ at slot $t$ according to Eqs. \eqref{eq12}-\eqref{eq15} given  $L_m$, $\kappa$, $P_{max}$, $\tau_0$ and $W$ and thus $B_m(t)$ is another local observation of VU $m$ at slot $t$ to reflect the stochastic task arrival and the uncertain channel condition.

\section{Problem Formulation}
\label{sec4}
In the system, statistics task arrival and uncertain channel condition are all unknown to each VU, thus we adopt DRL framework which includes state, action, policy and reward to formulate the power allocation problem in the VEC \cite{DRL}. Specifically, for each VU  at each slot $t$, VU observes the current local state $s_{t}$ and makes action $a_{t}$ based on $s_{t}$ according to policy $\mu$, i.e., the function that generates the action based on the state at each slot. Then VU receives a reward $r_{t}$ and observes the state at the next slot $s_{t+1}$, which is transited from the current state $s_{t}$. Next the state $s_{t}$, action $a_{t}$ and reward $r_{t}$ of VU at slot $t$ will be defined, respectively.

\subsection{State}
Different from the traditional centralized DRL-based offloading scheme in VEC, each VU observes its local state to determine the power allocation in this paper. Since the power consumption and delay are impacted by the stochastic task arrival and uncertain channel condition caused by the MIMO-NOMA channel interference and mobility of VUs, the local state should be selected to reflect the stochastic task arrival and uncertain channel condition as well as the mobility of the VU.

In the system model, the distance between VU $m$, and the antennas of the BS along $x$-axis at slot $t$, i.e., $d_m(t)$, determines the position of VU $m$ at slot $t$, which reflects the mobility of VU $m$. In addition, according to Eq. \eqref{eq11}, the SINR of VU $m$ at slot $t$, i.e., $\gamma_m(t-1)$, depends on ${\boldsymbol{g}_m^H(t-1)}$ that is related with the channel vector $\boldsymbol{h}_m(t-1)$, and thus $\gamma_m(t-1)$ can reflect the uncertain channel condition at slot $t$. Moreover, according to Eqs. \eqref{eq12}-\eqref{eq15}, the buffer length of VU $m$ at slot $t$, i.e., $B_m(t)$, is a function of $a_m(t-1)$ and $\gamma_m(t-1)$, where $a_m(t-1)$ reflects the stochastic task arrival and $\gamma_m(t-1)$ reflects the uncertain channel condition. Therefore, $B_m(t)$ reflects both stochastic task arrival and uncertain channel condition. As shown in the system model, $B_m(t)$, $\gamma_m(t-1)$ and $d_m(t)$ are all the local observations of VU $m$ at slot $t$, therefore the state of VU $m$ at slot $t$ can be defined as
\begin{equation}
s_{m,t}=[B_m(t),\gamma_m(t-1),d_m(t)].
\label{eq16}
\end{equation}
where $\gamma_m(t-1)$ depends on $\boldsymbol{h}_m(t-1)$ and $B_m(t)$ depends on $\gamma_m(t-1)$ and $a_m(t-1)$. Since $a_m(t-1)$ and $\boldsymbol{h}_m(t-1)$  are random values within continuous space, thus the state space of VU is continuous.

\subsection{Action}
Each VU $m$ allocates the local execution power and offloading power based on the local observed state $s_{m,t}$, thus the local execution and offloading power are defined as the action of VU $m$ at slot $t$, i.e.,

\begin{equation}
a_{m,t}=[p_{m,o}(t),p_{m,l}(t)].
\label{eq17}
\end{equation}
Note that similar to \cite{Kwak}, we consider the fine-grained computation applications, thus VU $m$ allocates the local execution and offloading power within continuous spaces in $[0,P_{m,l}]$ and $[0,P_{m,o}]$ to process the tasks, respectively. In this case the action space of VU $m$ is continuous.

\subsection{Reward function}

In this paper, VU $m$ aims to improve the network performance in terms of the power consumption and delay. As described in the computation model, the latency of task processing at the VEC server is negligible and the feedback delay during the offloading is also ignored at each slot. In this case the delay of task transmission is a constant, i.e., the duration of a slot. Thus the delay consumed by VU $m$ is impacted by the buffer delay that is proportional to the average buffer length according to the Little's Theorem \cite{Little_Th}. Therefore, the reward function of VU $m$ at slot $t$ is defined as
\begin{equation}
r_{m,t}= - \left[\omega_{1}\left(p_{m,o}(t)+p_{m,l}(t)\right)+\omega_{2}B_m(t)\right],
\label{eq18}
\end{equation}
where $\omega_{1}$ and $\omega_{2}$ are the nonnegative weighted factors.

The expected long-term discounted reward of VU $m$ is calculated as

\begin{equation}
J(\mu_m) \coloneqq \mathbb{E}_{\mu_m}\left[\sum_{t=1}^{N_j}\gamma^{t-1}r_{m,t}\right],
\label{eq19}
\end{equation}
where $\gamma \in [0,1]$ is the discounting factor and $N_j$ is the upper limit of slot index when VU $m$ moves on lane $j$. In this paper, we aim to find the optimal policy $\mu_m^*$ to maximize the expected long-term discounted reward of VU $m$.

Note that the network condition may be changed after each slot due to the dynamic VEC. At the beginning of each slot, each VU first observes its local state to acquire the changed network condition, then makes actions based on its own local observation. Therefore, the reliability of our proposed scheme can be guaranteed under the dynamic VEC network.

\section{Solution}
\label{sec5}

In this section, we first describe the training stage to obtain the optimal policy, then introduce the testing stage to test the performance under the optimal policy.

\subsection{Training stage}
Since the state and action spaces are continuous and the DDPG algorithm is suitable to solve the DRL-based problem under the continuous state and action space. Therefore, we utilize the DDPG algorithm to obtain the optimal policy in the training stage.

\begin{algorithm}
  \caption{Training Stage for the DDPG based Framework}
  \label{al1}
  \KwIn{$\gamma$, $\tau$, $\theta^m$, $\zeta^{m}$}
  \KwOut{optimized $\theta^{m*}$, $\zeta^{m*}$}
  Randomly initialize the $\theta^m$, $\zeta^{m}$\;
  Initialize target networks by $\zeta^{m'}\leftarrow\zeta^{m}$, $\theta^{m'}\leftarrow\theta^{m}$\;
  Initialize replay experience buffer $\mathcal{R}$\;
  \For{episode from $1$ to $K_{max}$ }
  {
    Reset simulation parameters for the VEC system model\;

    Receive initial observation state $s_{1}$\;
    \For{time slot $t$ from $1$ to $N_j$ }
    {
      Generate the power for local process and computation offloading according to the current policy and exploration noise $a_{m}=\mu_{\theta^m}(s_{m}|\theta^m)+\Delta_{t}$ \;
      Execute action $a_{m}$, observe reward $r_m$ and new state $s_m'$ from the system model\;
      Store transition $(s_{m},a_{m},r_{m},s_{m}')$ in $\mathcal{R}$\;
      \If {number of tuples in $\mathcal{R}$ is larger than $I$ }
      {
      Randomly sample a mini-batch of $I$ transitions tuples from $\mathcal{R}$\;
      Update the critic network by minimizing the loss function according to Eq. \eqref{eq23}\;
      Update the actor network according to Eq. \eqref{eq24}\;
      Update target networks according to Eqs. \eqref{eq25} and \eqref{eq26}.
      }
    }
  }
\end{algorithm}

The DDPG algorithm is based on actor-critic architecture. The actor is applied for policy improvement, and the critic is applied for policy evaluation. The DDPG algorithm adopts deep neural network (DNN) on actor and critic to efficiently approximate and evaluate the policy, respectively, thus forming the corresponding actor network and critic network. The actor network is used to approximate the policy $\mu_m$, where the approximated policy is denoted as $\mu_{\theta^m}$, the output of the actor network is the action based on the policy $\mu_{\theta^m}$ and observed state. In the DDPG algorithm, the optimal policy is obtained through iterative policy improvement and evaluation. Moreover, DDPG algorithm adopts the target networks including target actor network and target critic network to guarantee the stability of the algorithm. The architecture of target actor-network and target critic-network are the same with the actor network and critic network, respectively. The pseudocode of the proposed algorithm is described in Algorithm \ref{al1}. Let $\theta^{m}$ and $\zeta^{m}$ be the parameters of the actor network and critic network, respectively, and $\theta^{m'}$ and $\zeta^{m'}$ be the parameters of the target actor network and target critic network, respectively, $\Delta_t$ be the noise for action exploration at slot $t$. For ease of understanding, we further introduce the DDPG algorithm in detail as follows.

Firstly, $\theta^{m}$ and $\zeta^{m}$ are initialized randomly, while $\theta^{m'}$ and $\zeta^{m'}$ are initialized as $\theta^{m}$ and $\zeta^{m}$, respectively. A replay buffer $\mathcal{R}$ with sufficient space is constructed to cache transition at each slot (lines 1-3).

Then the algorithm is executed for $K_{max}$ episodes. In the first episode, the position of VU $m$ $(d_m(1),w_{m,j},0)$ is reset as the position that it enters the coverage area of the BS, i.e., $d_m(1)$ is set as $-\frac{D}{2}$, and $B_m(1)$ is initialized as half of the buffer size. Then $\boldsymbol{h_m}^s(0)$ is initialized randomly, and $\boldsymbol{g}_{m}(0)$ is calculated according to Eq. \eqref{eq7} based on $\boldsymbol{h_m}^s(0)$, given the $\boldsymbol{g}_m(0)$ the initial SINR $\gamma_m(0)$ is calculated according to Eq. \eqref{eq11}. Thus, VU $m$ can observe the state at slot $1$, i.e., $s_{m,1}=[B_m(1),\gamma_m(0),d_m(1)]$ (line 4-6).

Afterwards, the algorithm is executed iteratively from slot $1$ to slot $N_j$. Given the input $s_{m,1}$ the output of the actor network is $\mu_{\theta^m}(s_{m,1}|\theta^m)$. As a noise $\Delta_1$ is generated randomly and VU $m$ sets the action $a_{m,1}$ as $\mu_{\theta^m}(s_{m,1}|\theta^m)+\Delta_1$, thus the offloading power $p_{m,o}(1)$ and local execution power $p_{m,l}(1)$ are determined. Then, VU $m$ allocates offloading power and local execution power to process the task, while achieving the reward $r_{m,1}$ according to Eq. \eqref{eq13}. Then, BS adopts the ZF technology to determine the SINR $\gamma_m(1)$. Specifically, BS collects the channel vector of each vehicle, calculates $\boldsymbol{g}_{m}(1)$ according to Eq. \eqref{eq7}, and then determines the initial SINR $\gamma_m(1)$ according to Eq. \eqref{eq11} under the obtained $\boldsymbol{g}_m(1)$. Afterwards VU $m$ observes the next state $s_{m,2}=[B_m(2),\gamma_m(1),d_m(2)]$. Specifically, VU $m$ calculates $B_m(2)$ according to Eq. \eqref{eq8}, where $d_{m,l}(1)$ is calculated based on Eqs. \eqref{eq13}-\eqref{eq14} under $p_{m,l}(1)$ and  $d_{m,o}(1)$ is calculated according to Eq. \eqref{eq15} under $p_{m,o}(1)$. In addition, VU $m$ receives its SINR $\gamma_m(1)$ from BS. Moreover, VU $m$ calculates $d_m(2)$ according to Eq. \eqref{eq2} given the position $d_m(1)$. Then the tuple $(s_{m,1},a_{m,1},r_{m,1},s_{m,2})$ is stored in the replay buffer. When the number of tuples stored in the replay buffer is less than $I$, VU $m$ inputs the next state into the actor network and begins the next iteration (lines 7-10).

When the number of the stored tuples is larger than $I$, the parameters of actor network, critic network and target networks, i.e., $\theta^{m}$, $\zeta^{m}$, $\theta^{m'}$ and $\zeta^{m'}$, are updated literately to maximize $J(\mu_{\theta^m})$. The parameters of actor-network $\theta^{m}$ is updated with the policy gradient, i.e., updating $\theta^{m}$ toward the direction of the gradient of $J(\mu_{\theta^m})$, which is denoted as $\nabla_{\theta^{m}}J(\mu_{\theta^m})$. Let $Q^{\mu_{\theta^m}}(s_{m,t},a_{m,t})$ be the action-value function of VU $m$ following policy $\mu_{\theta^m}$ under $s_{m,t}$ and $a_{m,t}$, which stands for the expected discounted long-term reward of VU $m$ from slot $t$, i.e.,

\begin{equation}
Q^{\mu_{\theta^m}}(s_{m,t},a_{m,t})\coloneqq\mathbb{E}_{\mu_{\theta^m}}\left[\sum_{k=t}^{N_j}\gamma^{k-t}r_{m,k}\right].
\label{eq20}
\end{equation}
In \cite{D.silver}, Silver \emph{et al.} proved that solving $\nabla_{\theta^{m}}J(\mu_{\theta^m})$ can be substituted by solving the gradient of $Q^{\mu_{\theta^m}}(s_{m,t},a_{m,t})$, which is denoted as $\nabla_{\theta^{m}}Q^{\mu_{\theta^m}}(s_{m,t},a_{m,t})$. However, $Q^{\mu_{\theta^m}}(s_{m,t},a_{m,t})$ in Eq. \eqref{eq17} can not be calculated by Bellman equation due to the continuous action space \cite{Rich_Sutton}. To address this issue, the critic network adopts DNN parameterized by $\zeta^{m}$ to approximate the action-value function $Q^{\mu_{\theta^m}}(s_{m,t},a_{m,t})$, the action-value function approximated by critic-network is denoted as $Q^{\zeta^{m}}(s_{m,t},a_{m,t})$.

\begin{figure*}
\centering
\includegraphics[width=6in]{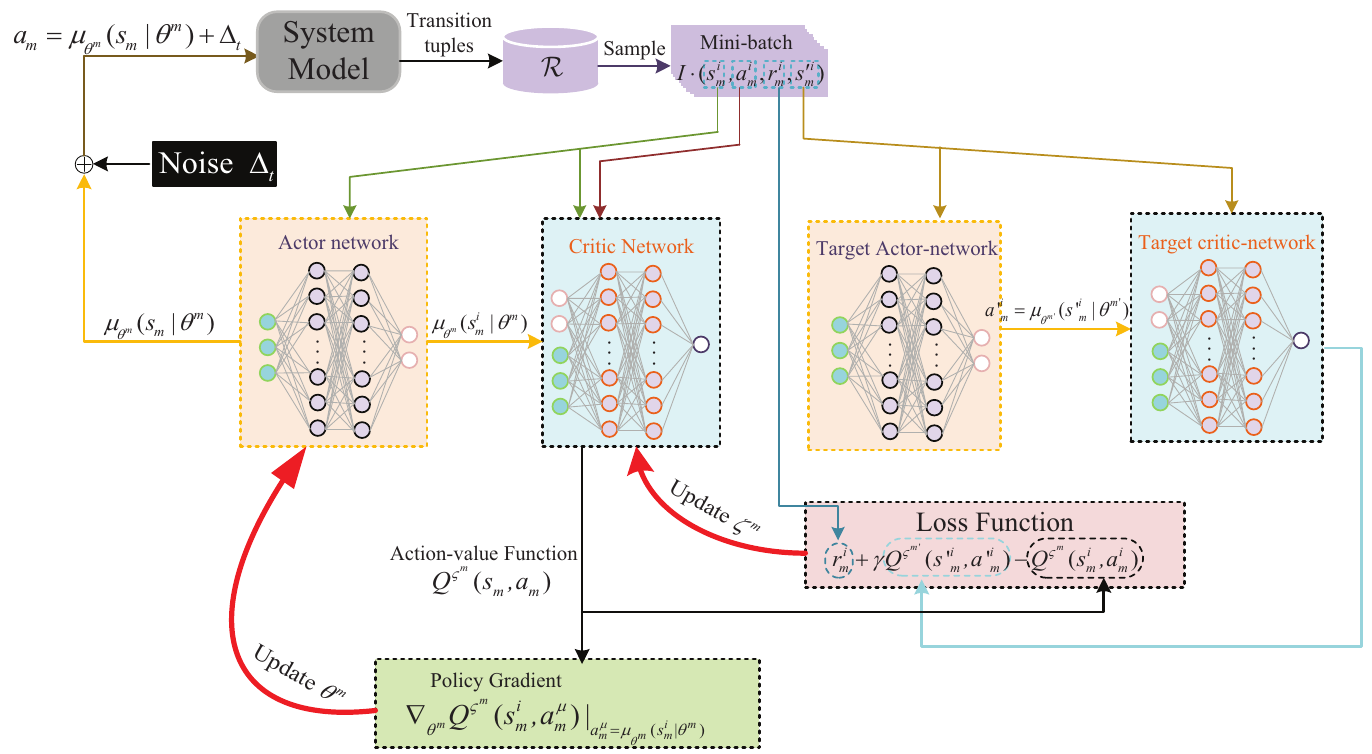}
\caption{Flow Diagram of DDPG}
\label{fig2}
\end{figure*}

The iteration in slot $t$ $(t=1,\ 2,\ \cdots,\ N_j)$ to update $\theta^{m}$, $\zeta^{m}$, $\theta^{m'}$ and $\zeta^{m'}$ is described as follows when the number of the stored tuples is larger than $I$. For simplicity, $r_{m,t}$, $s_{m,t}$, $a_{m,t}$, $s_{m,t+1}$ and $a_{m,t+1}$ are expressed as $r_{m}$, $s_m$, $a_m$, $s_{m}'$ and $a_{m}'$, respectively. VU $m$ first uniformly samples $I$ tuples from replay buffer to form a mini-batch. Let $(s_{m}^i,\ a_{m}^i,\ r_{m}^i,\ {s'}_{m}^{i})$ $(i=1,\ 2,\ \cdots,\ I)$ be the $i$-th tuple in the mini-batch. Then VU $m$ inputs each tuple into the target actor-network, target critic network and critic network. For tuple $i$, VU $m$ first inputs ${s'}_{m}^{i}$ into the target actor-network and outputs the action ${{a'}_{m}^{i}=\mu_{\theta^{m'}}({s'}_{m}^{i}|\theta^{m'})}$, then VU $m$ inputs ${s'}_{m}^{i}$ and $a_{m,i}'$ into the target critic-network and outputs the action-value function $Q^{\zeta^{m'}}({s'}_{m}^{i},{a'}_{m}^{i})$. After that VU $m$ calculates the target value as
\begin{equation}
y_{m}^{i}=r_{m}^{i}+\gamma Q^{\zeta^{m'}}({s'}_{m}^{i},{a'}_{m}^{i})|_{{a'}_{m}^{i}=\mu_{\theta^{m'}}({s'}_{m}^{i}|\theta^{m'})}.
\label{eq21}
\end{equation}

Then the loss function can be calculated as
\begin{equation}
\label{eq23}
L(\zeta^{m})=\frac{1}{I}\sum_{i=1}^{I}\left[y_{m}^{i}-Q^{\zeta^{m}}(s_{m}^{i},a_{m}^{i})\right]^2,
\end{equation}
and critic-network updates its parameters using $\nabla_{\zeta^{m}}L(\zeta^{m})$ to minimize the loss function through gradient descending\cite{DDPG}. (lines 11-13)

Similarly, actor-network updates its parameters using $\nabla_{\theta^m}J(\mu_{\theta^m})$ to maximize $J(\mu_{\theta^m})$ through gradient ascending  \cite{DDPG}, where $\nabla_{\theta^m}J(\mu_{\theta^m})$ is calculated by the action-value function which is approximated by critic-network, i.e., (line 14)
\begin{equation}
\begin{split}
&\nabla_{\theta^m}J(\mu_{\theta^m})\\
&\approx \frac{1}{I}\sum_{i=1}^{I}\nabla_{\theta^m}Q^{\zeta^{m}}(s_{m}^{i},a_{m}^{\mu})|_{a_{m}^{\mu}=\mu_{\theta^m}(s_{m}^{i}|\theta^m)}\\
&=\frac{1}{I}\sum_{i=1}^{I}\nabla_{a_{m}^{\mu}}Q^{\zeta^{m}}(s_{m}^{i},a_{m}^{\mu})|_{a_{m}^{\mu}=\mu_{\theta^m}(s_{m}^{i}|\theta^m)} \\&\quad\cdot \nabla_{\theta^m}\mu_{\theta^m}(s_{m}^{i}|\theta^m),
\end{split}
\label{eq24}
\end{equation}
here chain rule is applied since that $a_{m}^{\mu}=\mu_{\theta^m}(s_{m}^{i}|\theta^m)$ is the input of $Q^{\zeta^{m}}$.

At the end of slot $t$, VU $m$ updates the parameters of the target actor-network and target critic-network as
\begin{equation}
\label{eq25}
\zeta^{m'}\leftarrow \tau\zeta^{m} + (1-\tau)\zeta^{m'},
\end{equation}
\begin{equation}
\label{eq26}
\theta^{m^{\prime}}\leftarrow\tau\theta^{m}+(1-\tau)\theta^{m^{\prime}},
\end{equation}
where $\tau$ is a constant satisfying $\tau \ll 1$ (line 15).

Finally, VU $m$ input $s_m'$ into the actor network and begins the iteration in next slot. The episode is finished when the number of iterations reaches $N_j$. Then VU $m$ will initialize $B_m(1)$, $\gamma_m(0)$, $d_m(1)$ and start the next episode. The algorithm will finally terminate when the number of episodes reaches $K_{max}$, which means that the training stage is finished. The flow diagram of the DDPG algorithm is shown in Fig. \ref{fig2}.

\subsection{Testing stage}
The testing stage omits the critic network, target actor-network and target critic-network in the training stage and employs the optimal policy with optimized parameter $\theta^{m*}$ to test the performance. The pseudocode of the testing stage is shown in Algorithm \ref{al2}.
\begin{algorithm}
  \caption{Testing Stage for the DDPG based Framework}
  \label{al2}
  \For{episode from $1$ to $K'_{max}$ }
  {
    Reset simulation parameters for the VEC system model\;
    Receive initial observation state $s_{1}$\;
    \For{time slot $t$ from $1$ to $N_j$ }
    {
      Generate the power for local process and computation offloading according to the optimal policy $a_{m}=\mu_{\theta^m}(s_{m}|\theta^{m*})$ \;
      Execute action $a_{m}$, observe reward $r_m$ and new state $s_m'$ from the system model.
    }
  }
\end{algorithm}
\subsection{Complexity Analysis}
In this subsection, we analyze the complexity of the DDPG algorithm. Let  $G_A$ and $G_C$ be the computational complexity of computing gradients for actor network and critic network, respectively,  $U_A$ and $U_C$ be the computation complexity of updating parameters for actor network and critic network, respectively. Since the architecture of the target actor network and target critic network are the same as the actor network and critic network, the complexity of updating parameters for target networks are the same as actor network and critic network. The complexity of DDPG algorithm is affected by the number of slots for training. For each slot for training, the actor network and critic network compute gradients and update parameters, while the target networks update parameters without computing gradients. Thus the complexity of the DDPG algorithm in a slot for training is calculated as $O(G_A +G_C +2U_A +2U_C)$. In addition, the training and updating parameters process will not be activated until the tuples stored in replay buffer is larger than $I$, and the algorithm loops for Kmax episodes and each episode includes $N_j$ slots for training, thus the complexity of the DDPG algorithm is calculated as $O((K_{max}\cdot N_j-I)(G_A +G_C +2U_A +2U_C))$.

\section{Simulation Results and Analysis}
\label{sec6}
In this section, we conduct simulation experiments \footnote{The source code has been released at: https://github.com/qiongwu86/VEC\_DRL\_Doppler.git} to verify the effectiveness of the optimal power allocation scheme, i.e., the optimal policy, in the training and testing stage, respectively. The simulation tool is Python 3.6. The scenario is described in the system model. In the simulation experiments, both actor network and critic network are the four-layer fully connected DNN with two hidden layers which are equipped with $400$ and $300$ neurons, respectively. Adam optimization method \cite{adam} is adopted to update the parameters of critic network and actor network with learning rate as $10^{-3}$ and $10^{-4}$, respectively.

The noise $\Delta_{t}$ for exploration follows the Ornstein-Uhlenbeck (OU) process \cite{OUnoise} with the decay-rate  and variation as $0.15$ and $0.02$, respectively. The size of experience replay buffer is $|\mathcal{R}|$. The task arrivals at each slot follow Poisson distribution with mean task arrival rate $\lambda_m$. The maximum local process power $P_{m,l}$ is calculated according to Eq. \eqref{eq14} given the maximum allowable CPU frequency $F_{max}$. The small scale fading of VU $m$ is initialized as $\boldsymbol{h}_{m}^{s}(0) \sim \mathcal{CN}(\boldsymbol{0}, \boldsymbol{I}_N)$. The target VU, i.e., VU $m$, moves on lane $2$ with velocity $v_2$. And three other VUs on each of three lanes drive into the coverage of BS when $d_m(t)=0$. The remaining parameters and algorithm parameters are shown in TABLE \ref{tab2}.


\begin{table}
\caption{Values of the parameters in the experiments.}
\label{tab2}
\footnotesize
\centering
\begin{tabular}{|c|c|c|c|}
  \hline
  \multicolumn{4}{|c|}{Parameters of System Model}\\
  \hline
  \textbf{Parameter} &\textbf{Value} &\textbf{Parameter} &\textbf{Value}\\
  \hline
  $\sigma_R^2$ &$10^{-9}$ W &$h_r$ &$-30$ dB\\
  \hline
  $\Lambda$ &$7$ m & $W$ &$1$ MHz\\
  \hline
  $\tau_0$ &$20$ ms & $\kappa$ &$10^{-28}$\\
  \hline
  $v_1$ &20 m/s &$v_2$  &25 m/s\\
  \hline
  $v_3$ &30 m/s &$w$ &5 m\\
  \hline
  $L_m$ &$500$ cycles/bit & $\lambda_m$&$3$ Mbps\\
  \hline
  $H$ &10 m &N &4\\
  \hline
  $D$ &500 m &$P_{max,l}$ &1 W\\
  \hline
  $P_{max,o}$ &$1$ W &$F_{max}$ &$2.15$ GHz\\
  \hline
  $J$ &3 &$w_0$ &5 m \\
  \hline
  $T_s$ &$4$ s &$N_{max}$ &15\\
  \hline
  \hline
  \multicolumn{4}{|c|}{Parameters of DDPG}\\
  \hline
  \textbf{Parameter} &\textbf{Value} &\textbf{Parameter} &\textbf{Value}\\
  \hline
  $\omega_{1}$ &$0.9$ &$\omega_{2}$ &$0.1$\\
  \hline
  $\gamma$ &$0.99$ &$\tau$ &$0.001$\\
  \hline
  $K_{max}$ &$2000$ &$I$ &$64$\\
  \hline
  $K'_{max}$ &$10$ &$|\mathcal{R}|$ &$2.5\times10^5$\\
  \hline
\end{tabular}
\end{table}

\subsection{Training Stage}

\begin{figure}
\centering
\includegraphics[scale=0.5]{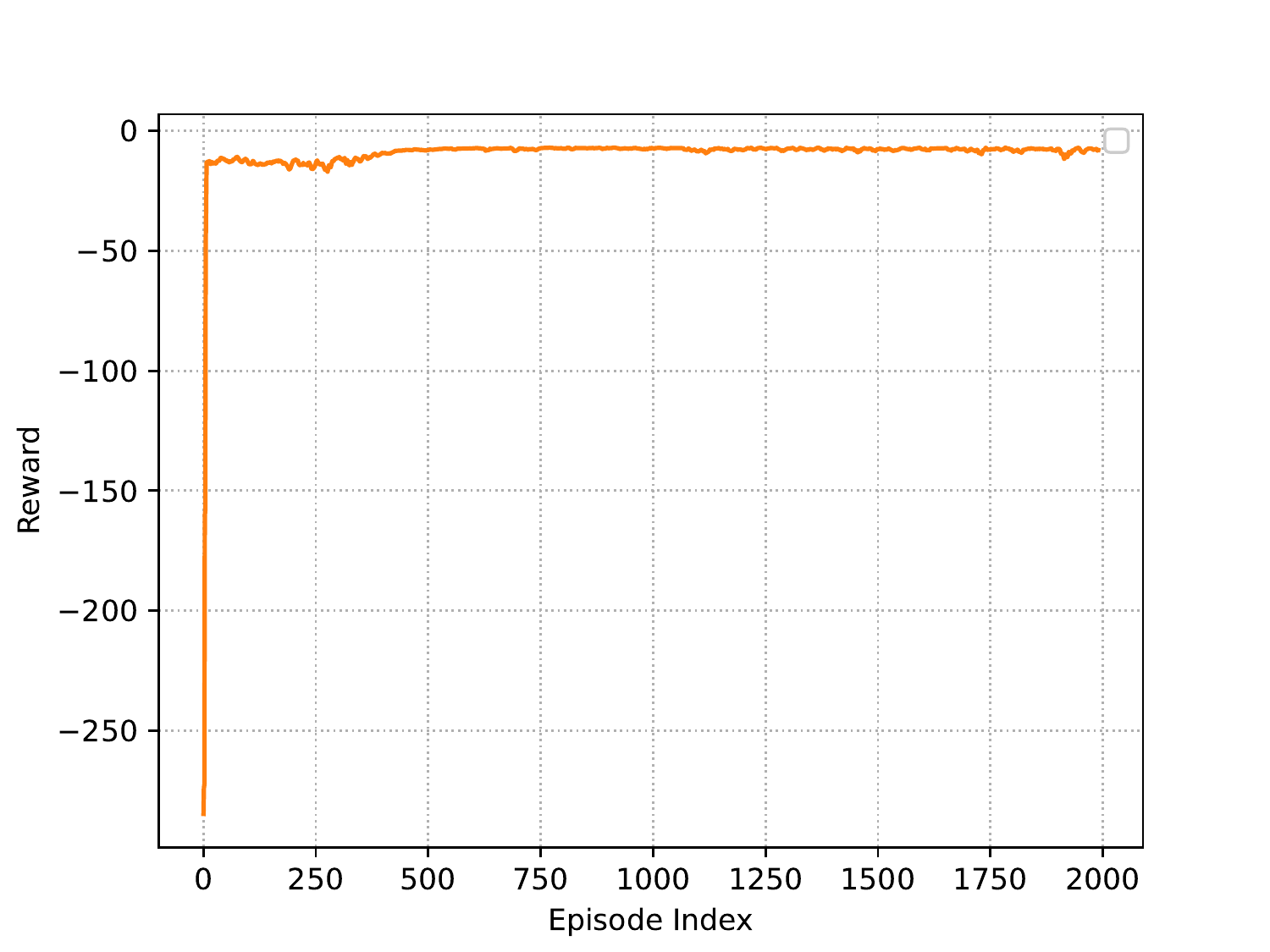}
\caption{Learning curve.}
\label{fig9}
\end{figure}

Fig.3 shows the learning curve of the training process, which reflects the average reward in each slot under different episodes. It can be seen that the average reward rises rapidly from episode $0$ to episode $10$, then the uptrend of the curve slows down from episode $10$ to $600$, which reflects that VU $m$ is learning the policy efficiently toward the optimal reward. Then the reward turns to be stable with little jitter, because the policy is adjusted slightly due to the exploration noise to prevent the policy from converging to local optimal value.

\subsection{Testing Stage}

In the testing stage, VU $m$ adopts learned policy in the training stage to test the performance. 
Figs. 4-6 compare the testing performance including the power consumption, buffer length and reward under the optimal policy with that under greedy local execution first (GD-Local) and greedy offload first (GD-Offload) policy, where the performance value is obtained through averaging the results obtained in $100$ episodes. GD-Local policy and GD-Offload policy are introduced as follow.

 \begin{itemize}
 \item GD-Local policy: VU $m$ firstly adopts the maximum local execution power to process tasks at each slot through local execution, while the remaining tasks are processed through offloading.

 \item GD-Offload policy: VU $m$ firstly adopts the maximum offloading power to process tasks through offloading at each slot, while the remaining tasks are processed through local execution.
\end{itemize}

\begin{figure*}
  \centering
    \subfigure[]{
    \begin{minipage}{5.6cm}
    \includegraphics[scale=0.4]{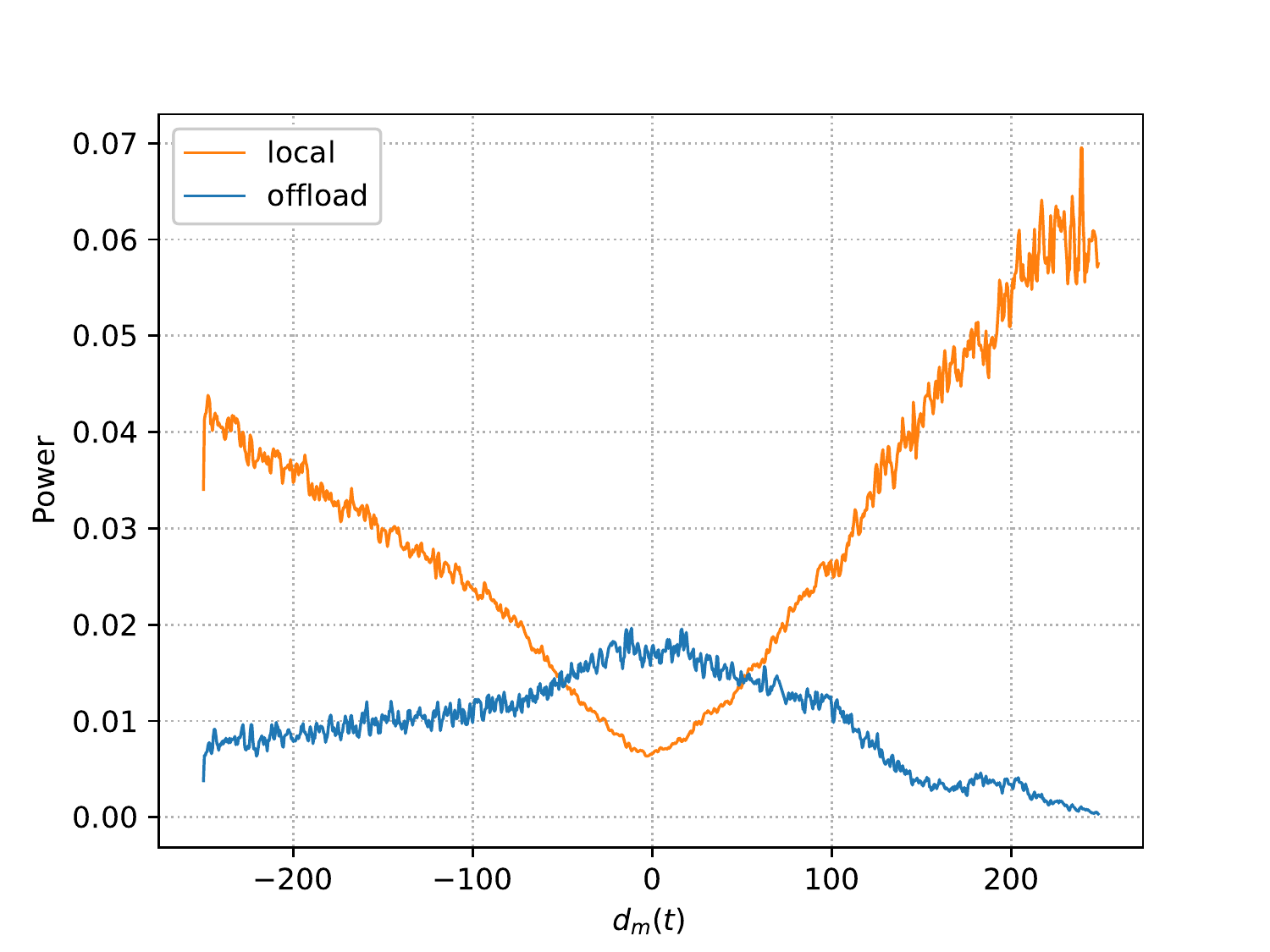}
    \end{minipage}}
  \subfigure[]{
    \begin{minipage}{5.6cm}
    \includegraphics[scale=0.4]{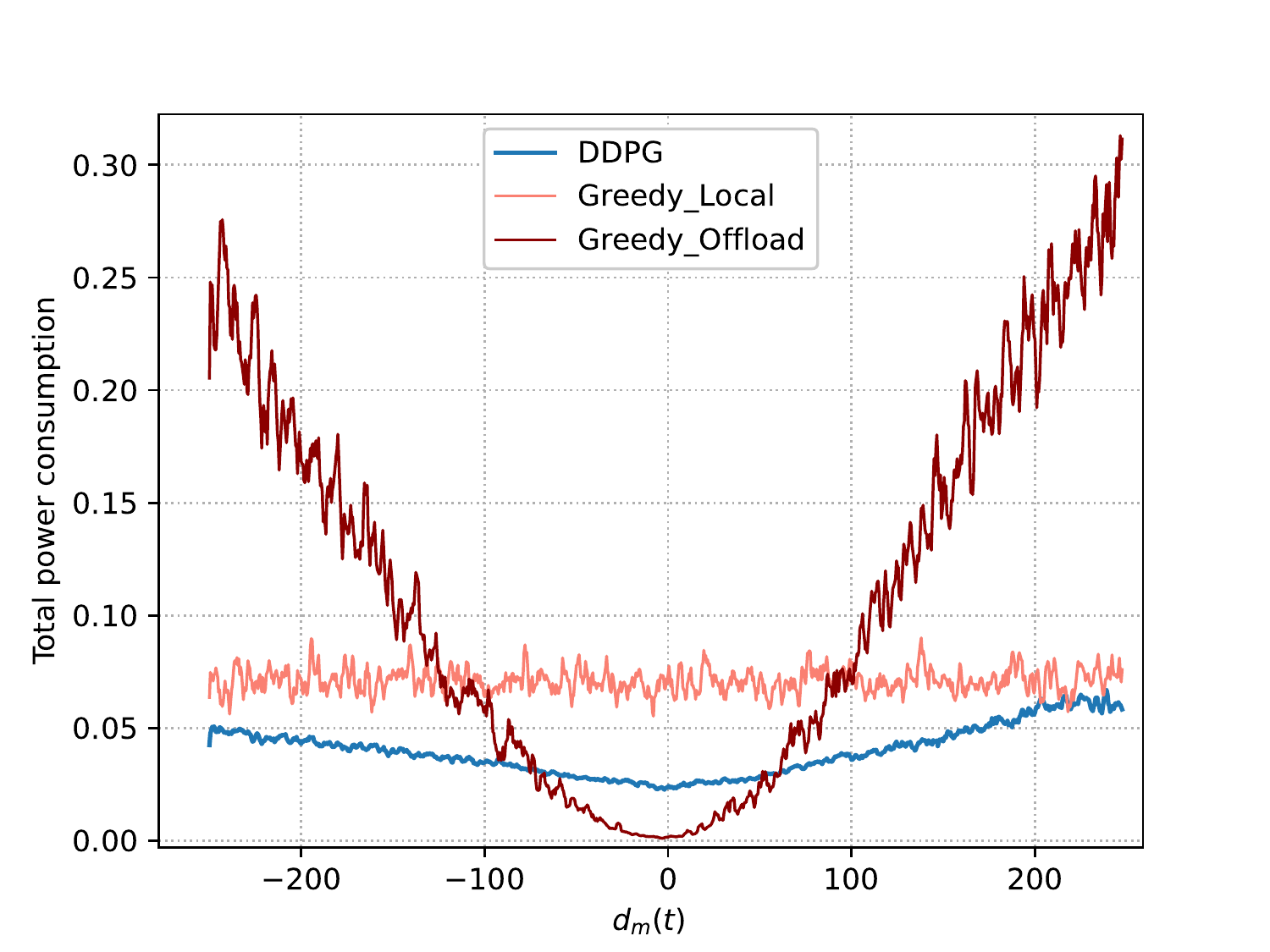}
    \end{minipage}}
    \caption{Power. (a) Optimal power allocation; (b) Total power consumption.}
    \label{power}
    \vspace{-0.4cm}
\end{figure*}

\begin{figure*}
  \centering
  \subfigure[]{
    \begin{minipage}{5.6cm}
    \includegraphics[scale=0.4]{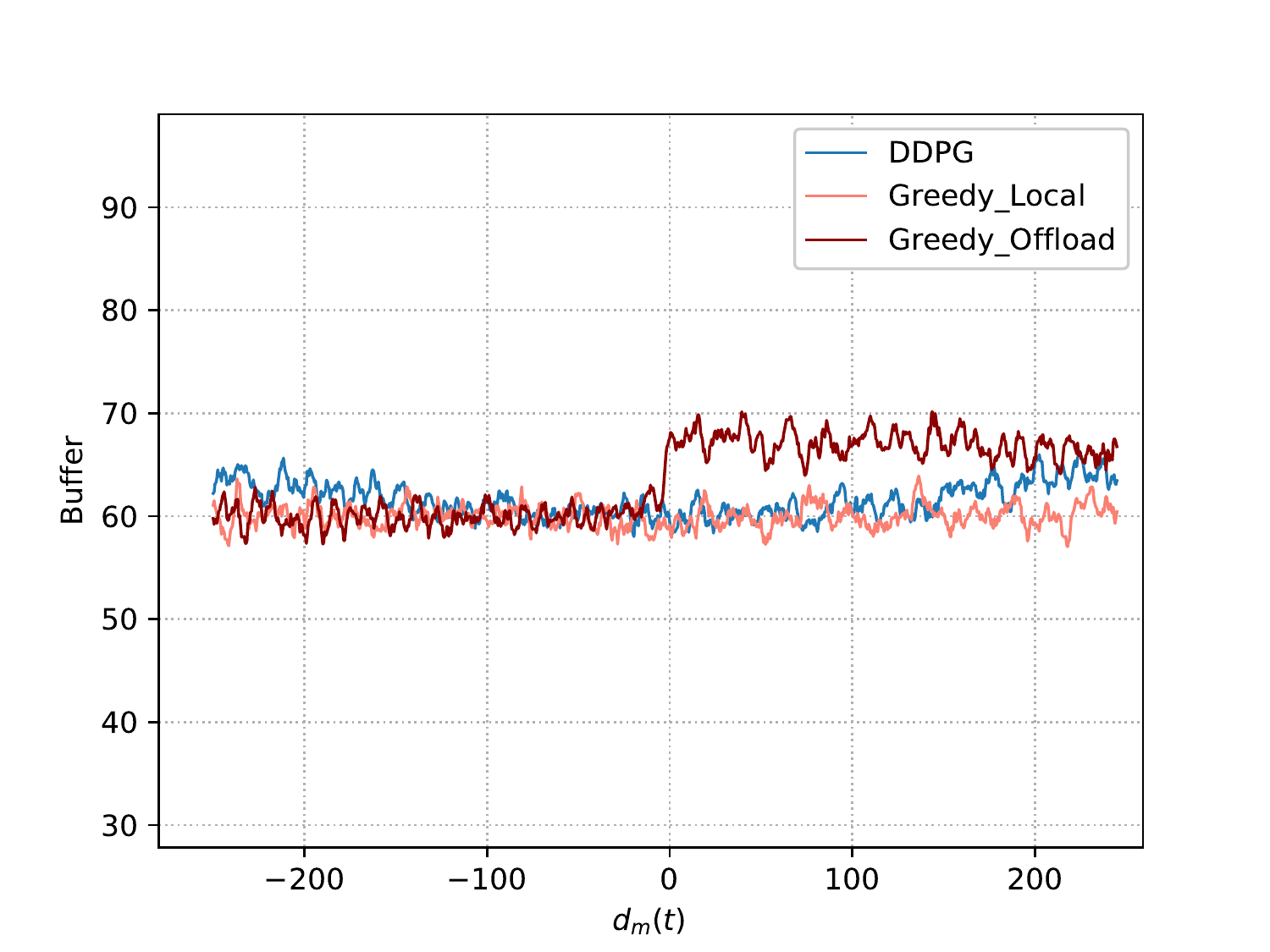}
    \end{minipage}}
  \subfigure[]{
    \begin{minipage}{5.6cm}
    \includegraphics[scale=0.4]{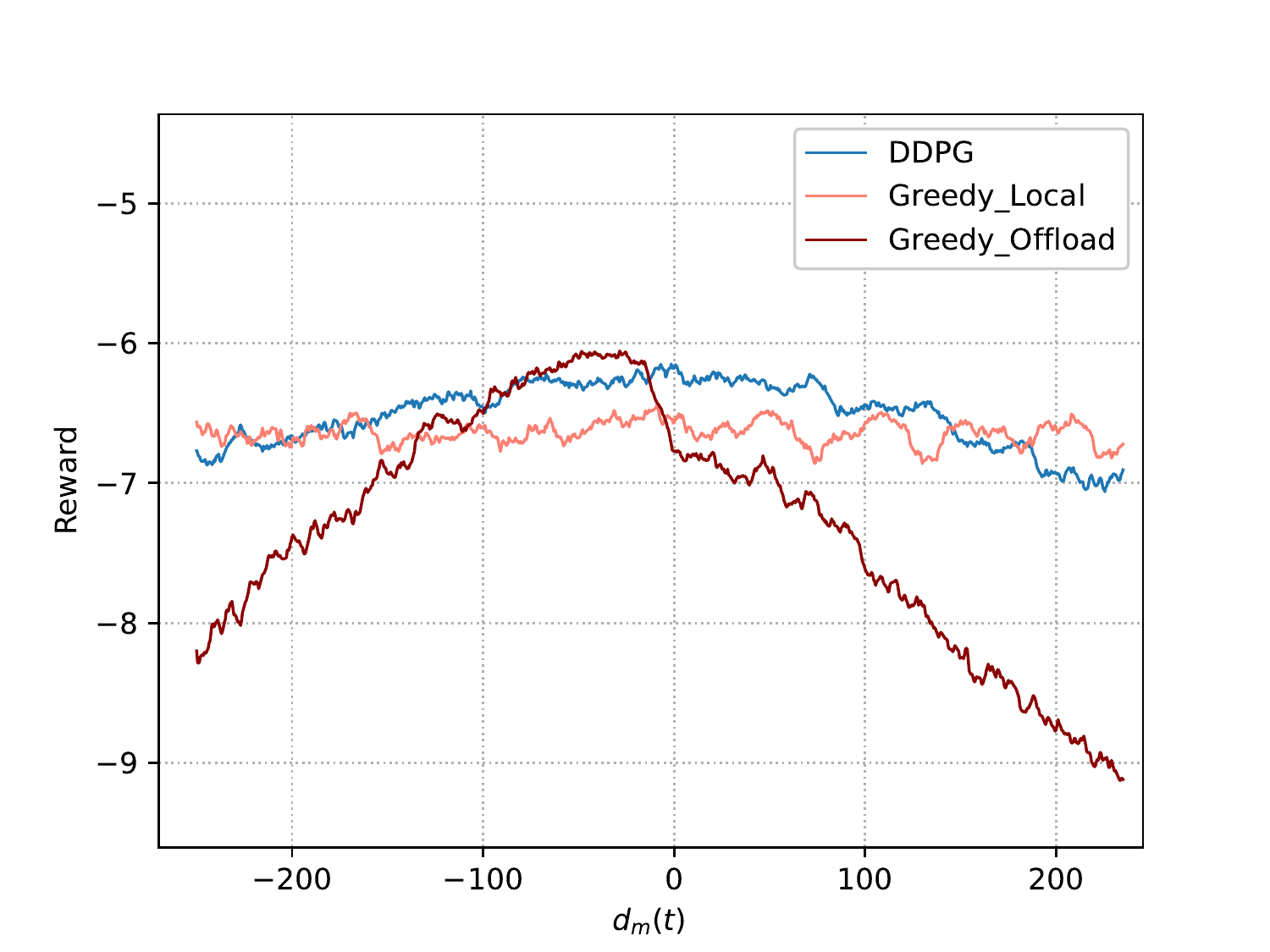}
    \end{minipage}}
    \caption{Performance. (a) Buffer length; (b) Reward.}
    \label{multi}
    \vspace{-0.4cm}
\end{figure*}

Fig. 4-(a) and (b) show the test results of power allocation and power comparison under three policies. Fig. \ref{power}-(a) compares the local execution power with offloading power under the optimal policy. It is seen that when $d_m(t)<0$, the local execution power decreases obviously and offloading power increases slowly. After that, the local execution power increases obviously and offloading power decreases slowly. It is because that according to Eq. \eqref{eq4}, the channel condition is impacted by path-loss. When VU $m$ is getting close to the BS, the path-loss is decreasing, thus leading to the better channel condition. Therefore, VU $m$ will consume more offloading power and less local execution power to process more tasks when it gets close to the BS. On the contrary, VU $m$ will consume less offloading power and more local execution power to process more tasks when it gets far away from the BS. The local execution power increases sharply after $d_m(t)>0$. This is because that three other vehicles drive into the BS's coverage area when $d_m(t)=0$, which imposes interference on VU $m$ and incurs the deteriorated channel condition. In this case, VU $m$ consumes more local execution power and less offloading power to process more tasks. 
Fig. \ref{power}-(b) shows the total power consumption under three policies. It can be seen that similar to the local execution power under the optimal policy in Fig. \ref{power}-(a), the total power under the optimal policy and GD-Offload decreases when $d_m(t)<0$, and increases when $d_m(t)>0$. This is because that the total power of the optimal policy is composed of both the local execution power and offloading power at each slot in Fig. \ref{power}-(a), where the local execution power overweighs offloading power. For the GD-Offload policy, the offloading power always keeps the maximum value, thus VU $m$ will process more tasks through offloading when it gets close to the BS, which results in less local execution power consumption in the GD-Offload policy. In contrast, it consumes more local execution power when the VU gets far away from the BS. Moreover, it can be seen that the total power almost does not change at each distance in the GD-Local policy. This is because that the local execution power always keeps the maximum power, while the offloading power is much smaller than the local execution power, which can be ignored in the total power under the GD-Local policy.

Figs. \ref{multi}-(a) and (b) compare the testing performances at each distance including buffer length and reward under three policies. 
Fig. \ref{multi}-(a) shows the buffer length under three policies. The buffer length of GD-Offload is increased when $d_m(t)=0$. This is because that more tasks cached in the buffer owing to the deteriorated channel condition. Moreover, the buffer length of the optimal policy also fluctuates around the mean of tasks arrival, which means that VU $m$ can process the tasks in time without increasing the buffer length when other VUs drive into the coverage of BS.
Figs. \ref{multi}-(b) compares the rewards of VU $m$ under the three policies. It can be seen that the reward of the optimal policy is usually larger than that of other two policies owing to the adaptive power allocation.

Fig. 6-(a) shows the average buffer length under three policies, where the average buffer length is obtained by averaging over the all slots in Fig. 5-(a). It can be seen that the average buffer lengths under three policies which are nearly equal to the mean of tasks arrival rate in each slot and do not change significantly. Fig. 6-(b) shows the average power consumption under three policies, where the average power consumption are obtained by averaging   over the all slots in Fig. 4-(a). It can be seen that compared with GD-Local, the average power consumption of the optimal policy is reduced by 44.2\%. Compared with GD-Offload, the average power consumption of the optimal policy is reduced by 63.0\%. 

\begin{figure*}
  \centering
  \subfigure[]{
    \begin{minipage}{5.6cm}
    \includegraphics[scale=0.4]{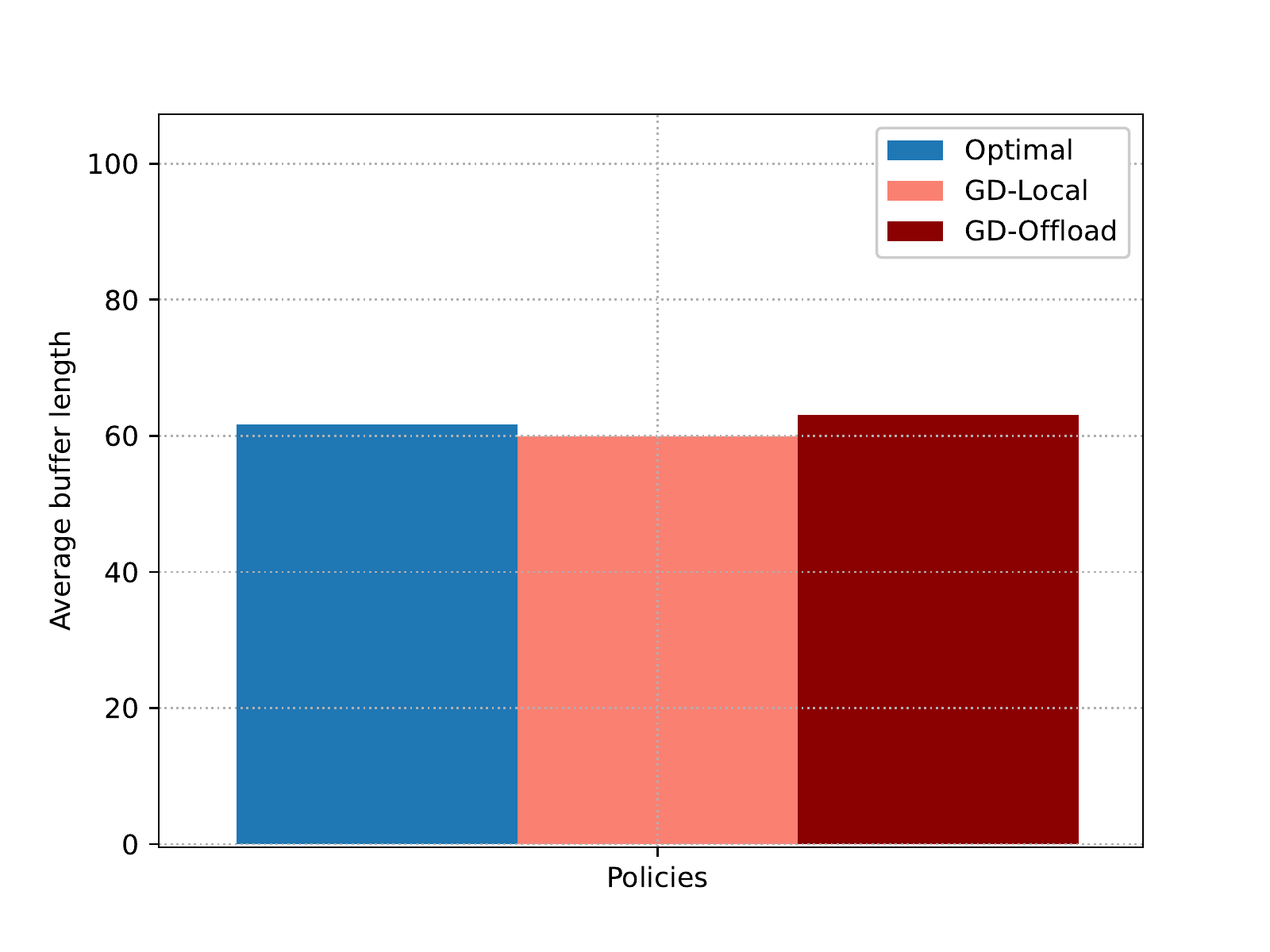}
    \end{minipage}}
  \subfigure[]{
    \begin{minipage}{5.6cm}
    \includegraphics[scale=0.4]{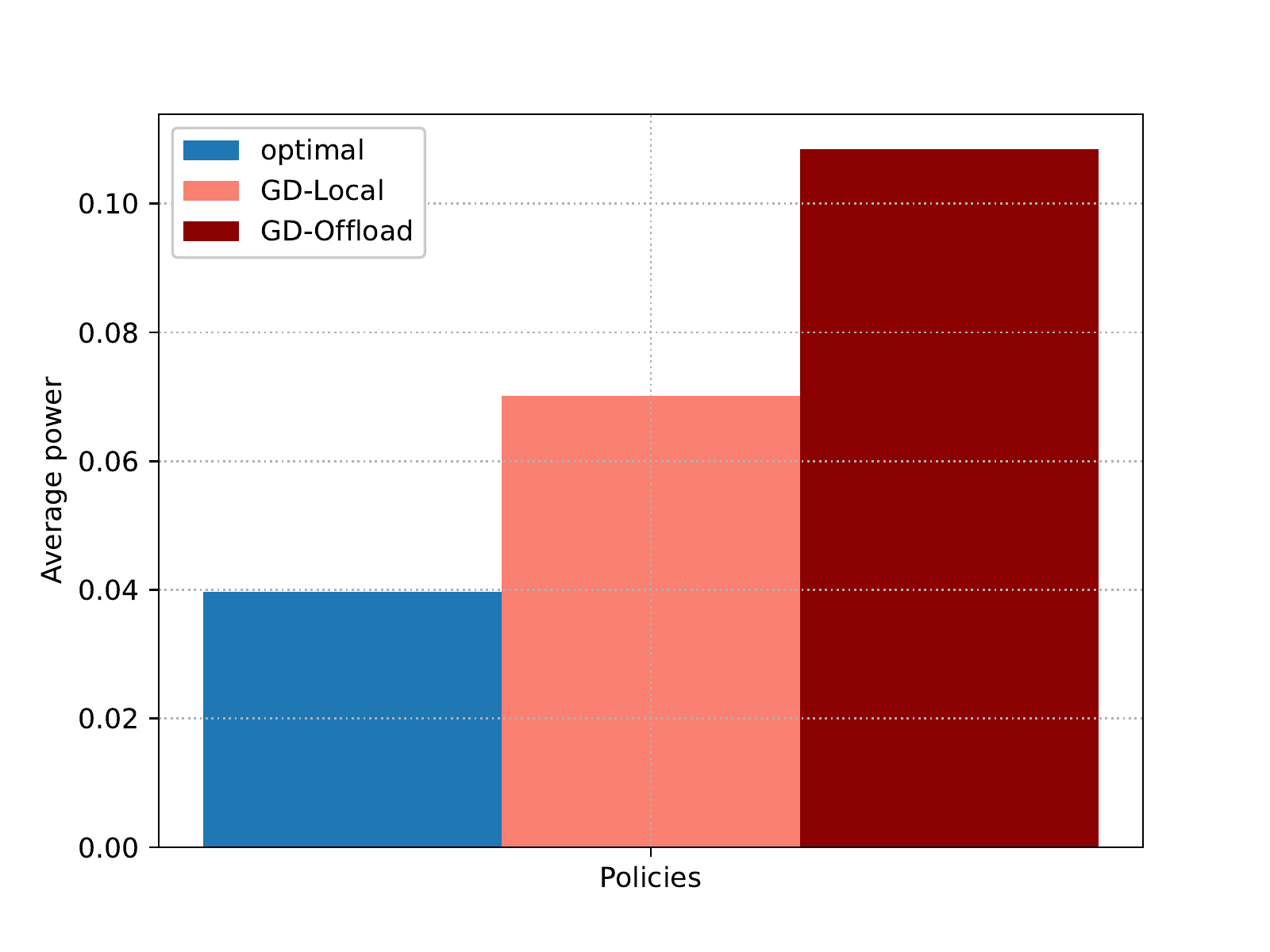}
    \end{minipage}}
  \subfigure[]{
    \begin{minipage}{5.6cm}
    \includegraphics[scale=0.4]{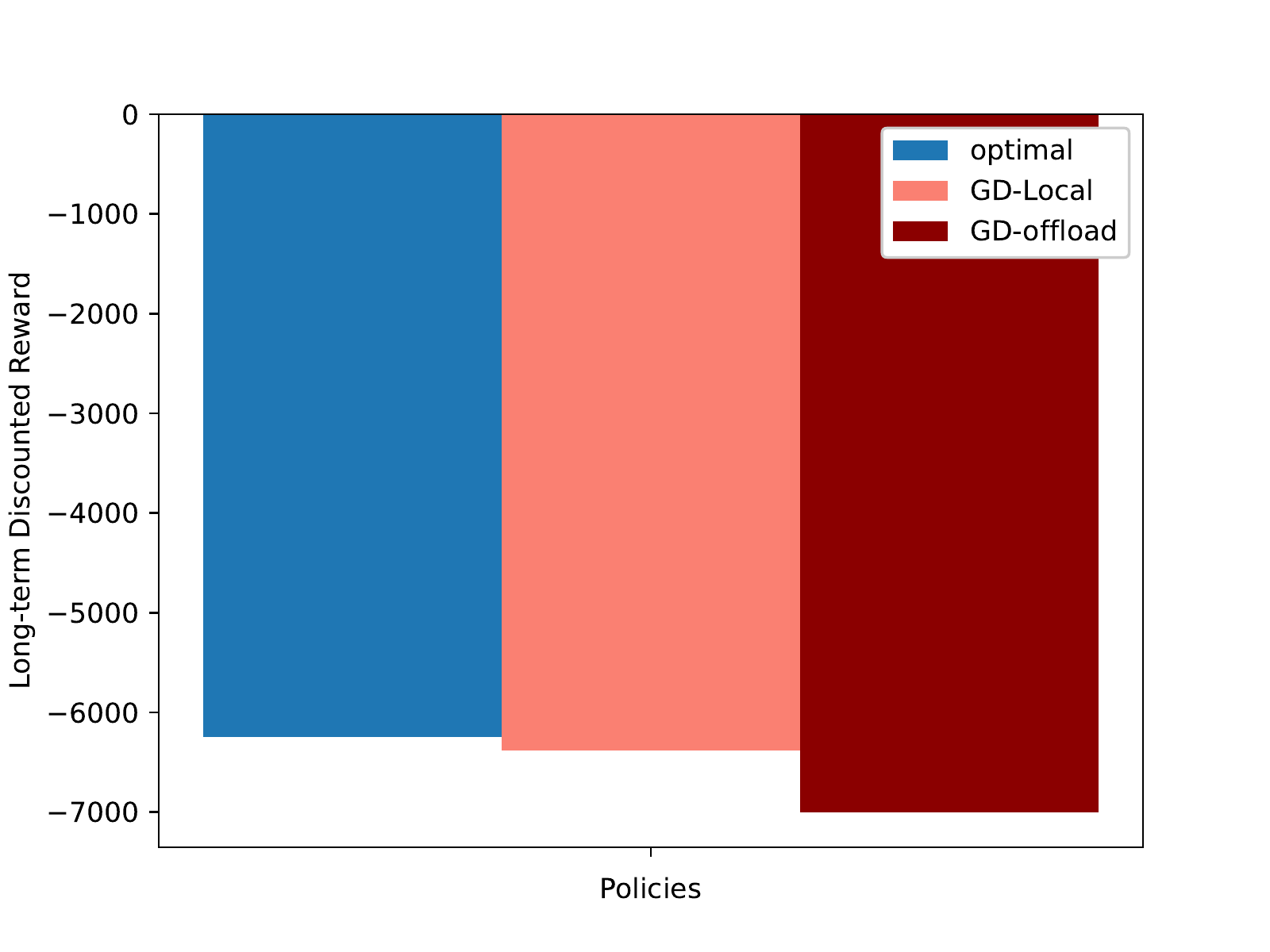}    
    \end{minipage}}
    \caption{(a) Average buffer length; (b) Average power consumption; (c) Long-term discounted reward.}
    \label{average}
    \vspace{-0.4cm}
\end{figure*}
\begin{figure*}
  \centering
  \subfigure[]{
    \begin{minipage}{5.6cm}
    \includegraphics[scale=0.399]{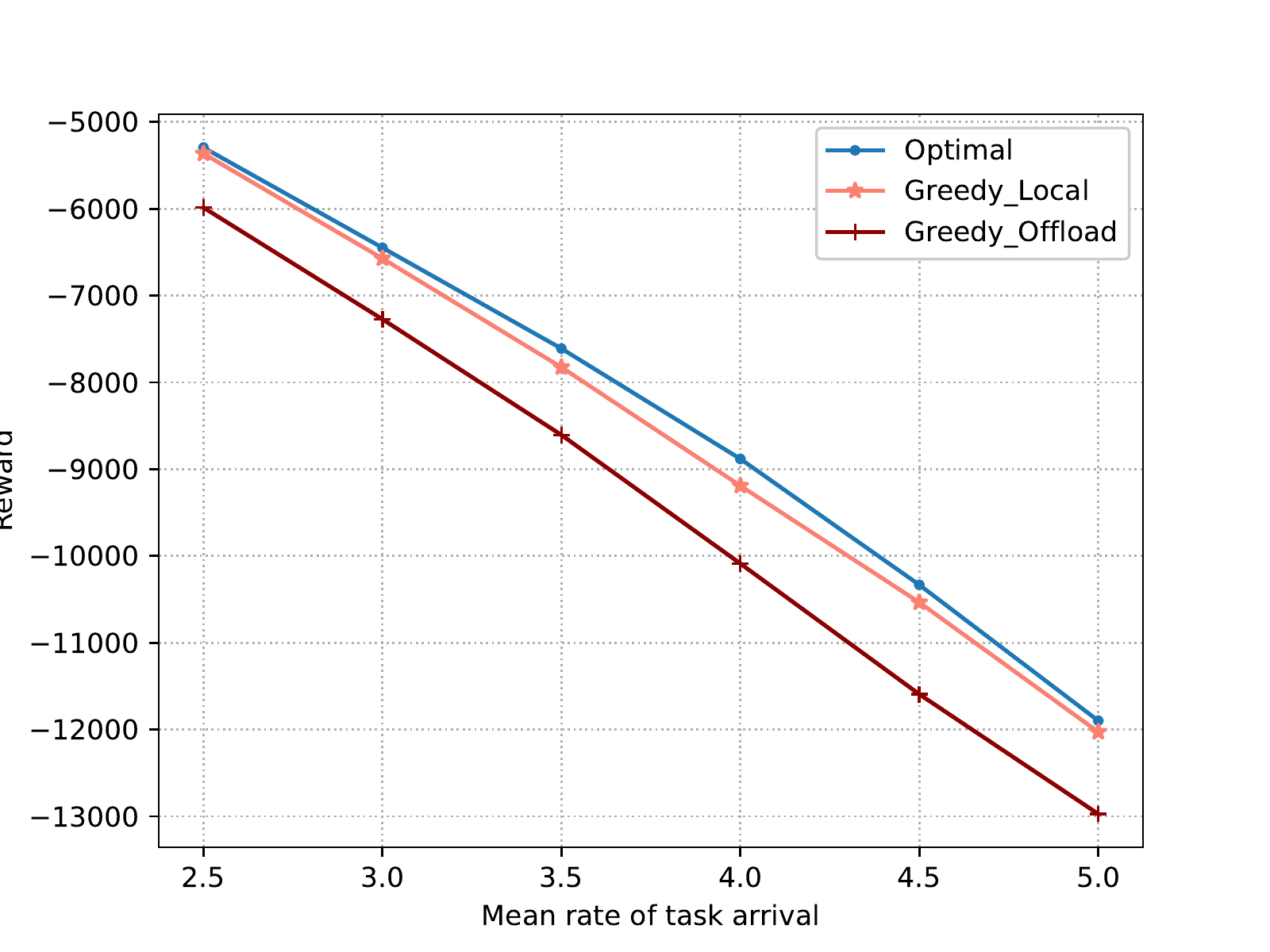}
    \end{minipage}}
  \subfigure[]{
    \begin{minipage}{5.6cm}
    \includegraphics[scale=0.399]{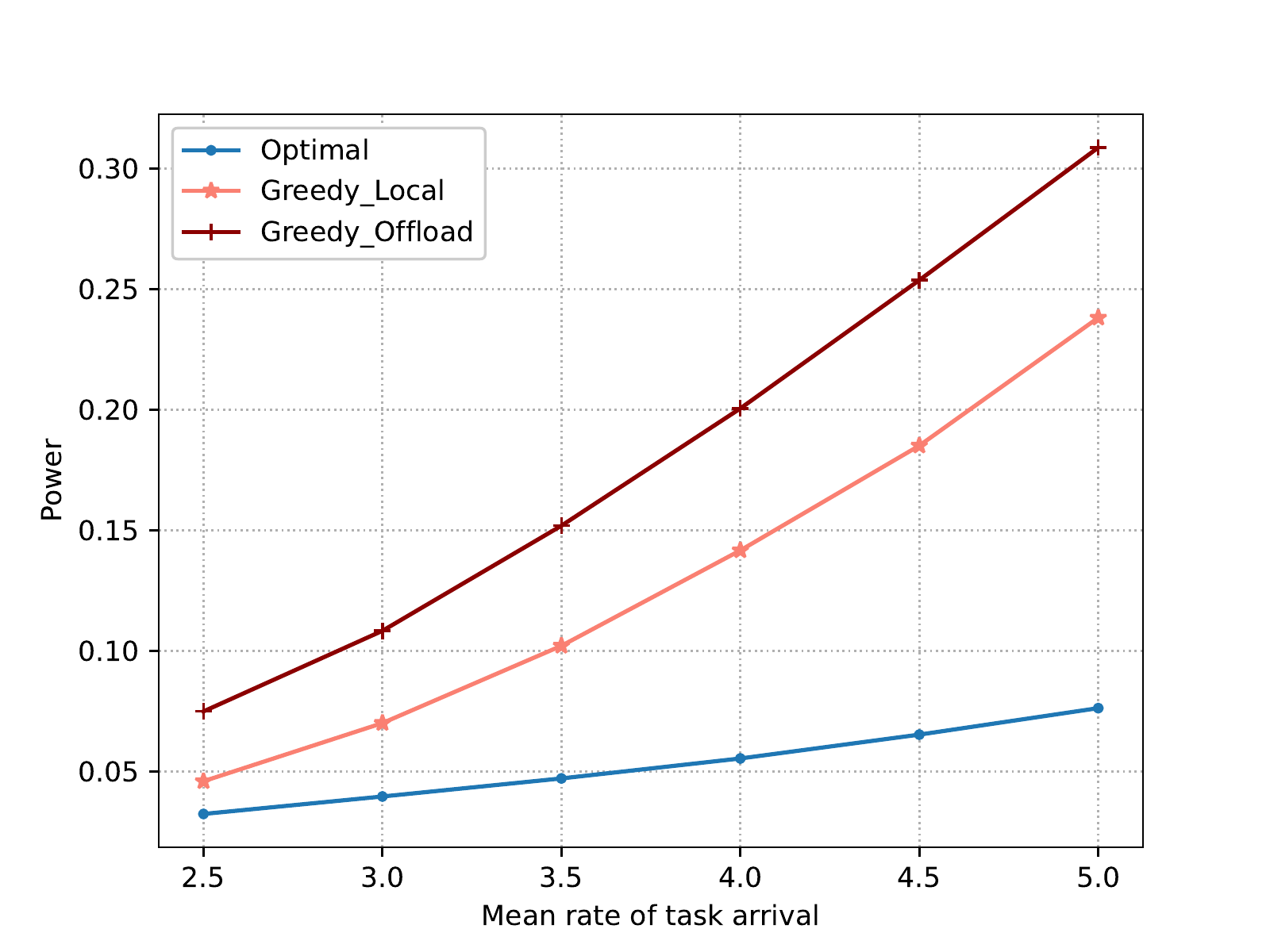}
    \end{minipage}}
  \subfigure[]{
    \begin{minipage}{5.5cm}
    \includegraphics[scale=0.4]{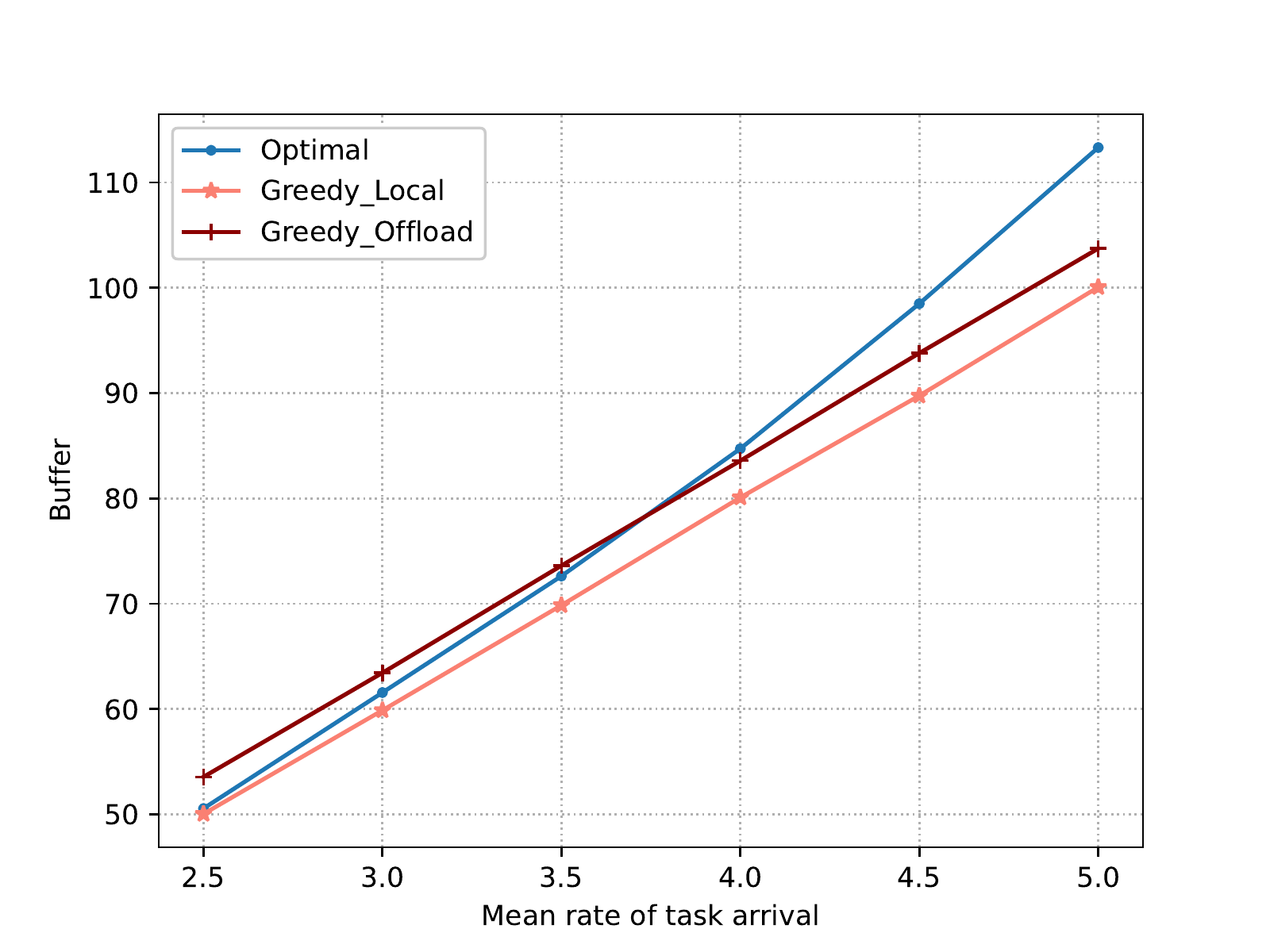}
    \end{minipage}}
    \caption{Performance vs task arrival. (a) Long-term discounted reward; (b) Power consumption; (c) Buffer length.}
    \label{tskarivl}
    \vspace{-0.4cm}
\end{figure*}
Fig. 6-(c) compare the long-term discounted reward under the three policies. As one can see, the optimal policy always has a higher the long-term discounted reward than other policies. This is because the optimal policy can adaptively adjust power allocation to maximize the long-term discounted reward.

Figs. \ref{tskarivl}-(a), (b) and (c) illustrate the long-term discounted reward, power consumption and buffer length of the three policies under different task arrivals, respectively. It can be seen that the long-term discounted rewards of the three policies decrease as the task arrival rate increases. As seen, increasing task arrival will lead to more power consumption and longer buffer length, thus degrading the reward according to Eq. \eqref{eq18}. It also can be seen that the optimal policy outperforms GD-Local and GD-Offload policies in terms of power consumption and long-term discounted reward, but it has a slightly higher buffer length than other policies. This is because the objective of the optimal policy is to maximize the long-term discounted reward by making a tradeoff between power the consumption and buffer length, which may lead to the compromise for buffer length. 
\section{Conclusions}
\label{sec7}
In this paper, we considered the stochastic task arrival and uncertain channel condition caused by both the MIMO-NOMA channel interference and VU mobility in VEC, and proposed a decentralized power allocation scheme based on the DRL to maximize the long-term reward including the power consumption and delay. We first formulated the system model and then constructed a DRL framework where the state is defined as the local observations. The DDPG algorithm has been adopted to learn the optimal policy. Extensive simulations have demonstrated the optimal policy outperforms the other existing policies. According to the theoretical analysis and simulation results, the conclusions can be made as follows:

\begin{itemize}
\item Since the channel condition is impacted by the mobility of VU, more offloading power and less local execution power are consumed to process more tasks when the VU gets close to the BS. On the contrary, less offloading power and more local execution power are consumed when it gets far away from the BS.

\item  In order to maximize the long-term discounted reward, the optimal policy emphasizes the power consumption and compromises the buffer length as compared to the GD-Local and GD-Offload policies given different tasks arrival rates.

\end{itemize}

For the future work, we will consider the data freshness  to design the offloading scheme in VEC.

\ifCLASSOPTIONcaptionsoff
  \newpage
\fi

\bibliographystyle{IEEEtran}
\bibliography{IEEEabrv,mybibfile}

\end{document}